\documentclass{aa}
\usepackage{graphicx}
\usepackage{natbib}
\bibpunct{(}{)}{;}{a}{}{,}
\usepackage{txfonts}

\newcommand{\ha}{H$\alpha$}
\newcommand{\hb}{H$\beta$}

\newcommand{\kms}{\,km\,s$^{-1}$}  
\newcommand{\myr}{\,$M_{\sun}\,{\rm yr}^{-1}$}
\newcommand{\ro}{\,$R_{\sun}$}
\newcommand{\mo}{\,$M_{\sun}$}
\newcommand{\lo}{\,$L_{\sun}$}
\newcommand{\cmt}{\,cm$^{-3}$}

\newcommand{\cmd}{\,cm$^{-2}$}
\newcommand{\cmdd}{\,cm$^{2}$}
\newcommand{\es}{$\rm\,erg\,s^{-1}$}
\newcommand{\ecs}{$\rm\,erg\,cm^{-2}\,s^{-1}$}

\begin{document}

\title{New outburst of the symbiotic nova AG~Peg after 165 years}

\author{A.~Skopal          \inst{1}
     \and S.~Yu.~Shugarov  \inst{1,2} 
     \and M.~Seker\'a\v{s} \inst{1}   
     \and M.~Wolf          \inst{3}   
     \and T.~N.~Tarasova   \inst{4}   
     \and F.~Teyssier      \inst{5}   
     \and M.~Fujii         \inst{6}   
     \and J.~Guarro        \inst{7}   
     \and O.~Garde         \inst{8}   
     \and K.~Graham        \inst{9}   
     \and T.~Lester        \inst{10}   
     \and V.~Bouttard      \inst{11}  
     \and T.~Lemoult       \inst{12}  
     \and U.~Sollecchia    \inst{13}  
     \and J.~Montier       \inst{14}  
     \and D.~Boyd          \inst{15}  
}
\institute{Astronomical Institute, Slovak Academy of Sciences, 
           059\,60 Tatransk\'{a} Lomnica, Slovakia 
\and
Sternberg Astronomical Institute, Moscow State University, 
Universitetskij pr., 13, Moscow, 119991, Russia
\and
Astronomical Institute, Charles University Prague, CZ-180\,00
Praha 8, \mbox{V Hole\v{s}ovi\v{c}k\'ach} 2, The Czech Republic
\and
Crimean Astrophysical Observatory, 298409 Nauchny, Crimea, Russia
\and
67 Rue Jacques Daviel, Rouen 76100, France
\and
Fujii Kurosaki Observatory, 4500 Kurosaki, Tamashima, Kurashiki, 
Okayama 713-8126, Japan 
\and
Balmes 2, 08784 PIERA, Barcelona, Spain
\and 
Observatoire de la Tourbi\`ere, 38690 Chabons, France
\and
23746 Schoolhouse Road, Manhattan, Illinois, USA 60442
\and
1178 Mill Ridge Road, Arnprior, ON, K7S3G8, Canada
\and
21 rue de Gu\'emar, 68000 Colmar, France
\and
Chelles Observatory, 23 avenue h\'enin, 77500 Chelles, France
\and
Via dei Malatesta 10, 67100 L'Aquila, Italy
\and
30 rue de la Boulais - 35000 Rennes, France
\and
West Challow Observatory, Oxfordshire OX12 9TX, UK
}
\date{Received / Accepted}

\abstract
 {
AG~Peg is known as the slowest symbiotic nova, which 
experienced its nova-like outburst around 1850. After 165 
years, during June of 2015, it erupted again showing 
characteristics of the Z~And-type outburst. 
 }
 {
The primary objective is to determine basic characteristics, 
the nature and type of the 2015 outburst of AG~Peg. 
 }
 {
We achieved this aim by modelling the spectral energy distribution 
using low-resolution spectroscopy (330--750\,nm; R = 500--1000), 
medium-resolution spectroscopy (420--720\,nm; $R\sim 11000$), and 
$UBVR_{\rm C}I_{\rm C}$ photometry covering the 2015 outburst 
with a high cadence. Optical observations were complemented with 
the archival \textsl{HST} and \textsl{FUSE} spectra from 
the preceding quiescence. 
 }
 {
During the outburst, the luminosity of the hot component was 
in the range of 2--11$\times 10^{37}\,(d/0.8\,{\rm kpc})^2$\es, 
being in correlation with the light curve (\textsl{LC}) profile. 
To generate 
the maximum luminosity by the hydrogen burning, the white dwarf 
(WD) had to accrete at $\sim 3\times 10^{-7}$\myr, which exceeds 
the stable-burning limit and thus led to blowing optically thick 
wind from the WD. We determined its mass-loss rate to a few 
$\times 10^{-6}$\myr. At the high temperature of the ionising 
source, $1.5-2.3\times 10^5$\,K, the wind converted a fraction 
of the WD's photospheric radiation into the nebular emission 
that dominated the optical. A one order of magnitude increase of 
the emission measure, from 
a few $\times 10^{59}\,(d/0.8\,{\rm kpc})^2$\cmt\ during 
quiescence, to a few $\times 10^{60}\,(d/0.8\,{\rm kpc})^2$\cmt\ 
during the outburst, caused a 2\,mag brightening in the 
\textsl{LC}, which is classified as the Z~And-type of the 
outburst. 
 }
 {
The very high nebular emission and the presence of a disk-like 
\ion{H}{i} region encompassing the WD, as indicated by a 
significant broadening and high flux of the Raman-scattered 
\ion{O}{vi} 6825\,\AA\ line during the outburst, is consistent 
with the ionisation structure of hot components in symbiotic 
stars during active phases. 
}
\maketitle
%
%
\section{Introduction}
\label{sec:intr}
Symbiotic stars are the widest interacting binary systems 
comprising a cool giant as the donor star and a hot compact star, 
mostly a white dwarf (WD), 
accreting from the giant's wind 
\citep[][]{ms99}. Their orbital periods run from hundreds of days 
(S-type systems containing a normal giant) to a few times 
10--100 years (D-type systems containing a Mira variable 
surrounded by a dust shell). 
The accreting WD represents a strong source of the extreme 
ultraviolet radiation ($T_{\rm h} \ga 10^5$\,K, 
$L_{\rm h} \sim 10^1-10^4$\lo) in the binary that ionises 
a fraction of the wind from the giant giving rise to the nebular 
emission \citep[e.g.][ hereafter STB]{stb}. 
This configuration represents the so-called {\em quiescent phase}, 
during which the symbiotic system releases its energy approximately 
at a constant rate and temperature. 
The observed luminosities of accreting WDs are powered by two 
main sources of the energy. In rare cases, low luminosities of 
$\sim 10^1 - 10^2$\lo\ are generated solely by the accretion 
process onto the WD 
\citep[e.g. EG~And, 4~Dra, SU~Lyn, see][]{sk05a,mukai+16}, 
when its gravitational potential energy is converted into the 
radiation by the disk. 
In most cases, high luminosities of a few times $10^3$\lo\ 
\citep[e.g.][]{mu+91,sk05b} are generated by stable nuclear 
hydrogen burning on the WD surface, which requires a certain 
range of accretion rates \citep[][]{tutyun76,paczyt78}. 

Sometimes, the symbiotic system changes its radiation 
significantly, brightens up in the optical by a few magnitudes 
and shows signatures of a mass-outflow. We name this stage as 
the {\em active phase}. 
According to the brightness variation we distinguish two 
subclasses: (i) The `Z~And-type' outbursts with amplitudes of 
1--3\,mag in the optical, evolving on the timescale of weeks 
to years. 
This type of outburst can result from an increase in the accretion 
rate above that sustaining the stable burning, which leads to 
expansion of the burning envelope simulating an A--F type 
pseudophotosphere \citep[][]{tutyun76,pacrud80}. 
(ii) The nova-like outbursts with amplitudes $> 3$\,mag that, 
after prolonged accretion by the WD from the giant's wind, undergo 
a thermonuclear outburst. Depending on the WD mass and the 
accretion rate we observe either very slow novae, whose outbursts 
last for dozens of years, or very fast novae lasting for several 
days to months and recurring during several years to decades 
\citep[see modelling by][]{livio+89,hk01,yaron+05}. The former 
are called `symbiotic novae' whereas the latter are known as 
`symbiotic recurrent novae'. Their properties have been discussed 
by many authors \citep[e.g.][]{allen80,mn94,munari97,mika11}. 

AG~Peg is a symbiotic binary comprising a M3\,III giant 
\citep[][]{kfc87} and a WD on an 818-d orbit 
\citep[e.g.][]{fekel+00}. This is a bright ($V \approx 8.5$), 
and thus a well studied object. Measuring its light variations 
from before 1850 \citep[see Fig.~1 of][]{boyar67} allowed for 
its classification as a symbiotic nova \citep[][]{allen80}. 
The outburst began in 1850 \citep[][]{lundmark21} when 
it rose in brightness from $\sim 9$ to reach a maximum of 
$\sim 6$\,mag around 1885. Afterwards, AG~Peg followed a gradual 
decline to around 2000, and then, varying between 8.5 and 9.0\,mag 
in $V$, kept its brightness until June 2015. Such an evolution 
of the \textsl{LC} together with other spectrophotometric 
parameters represents the slowest nova outburst ever recorded 
\citep[][]{kenyon+93}. 

The evolution of luminosity and temperature of the burning WD 
during the whole outburst was described by \cite{mn94} and 
\cite{altamore+97}. 
The peak luminosity, reached during 1940-50, was followed with 
a gradual decline to 1993 along a slowly increasing temperature. 
The following study by \cite{kenyon+01} confirmed the continuing 
slow decline of the AG~Peg luminosity by a factor of 2--3 from 
1980--1985 to 1997. 

Similar behaviour was indicated in the mass-loss rate from the 
hot component, $\dot M_{\rm h}$. Based on radio observations 
performed during 1973--75, \cite{gk77} estimated 
$\dot M_{\rm h} = 10^{-6}$\myr. Using the radio flux at 5\,GHz 
from 1984--87, \cite{kenny+91} estimated an upper limit of 
$\dot M_{\rm h}$ to $1.1\times 10^{-6}$\myr. 
During 1978--1993, a decline in $\dot M_{\rm h}$ from 
$\sim 10^{-6}$ to $(2-3)\times 10^{-7}$\myr\ was suggested by 
\cite{kenyon+93}. 

The presence of strong winds from both the binary components 
was often used to identify main emission regions in AG~Peg 
within the colliding winds model \citep[e.g.][]{tomov93,
contini97,contini03,tomov+01,eriksson+04,kenny+07}. 
The most direct indication of colliding winds in AG~Peg is 
the detection of X-ray emission from a very hot, optically 
thin plasma \citep[][]{muerset+95}. 
According to the gradual decrease of the hot component luminosity, 
\cite{zt95} predicted that the colliding-wind stage would end 
between 1999 and 2004, and accretion from the wind would 
recommence. 
From around 1997 to June 2015, AG~Peg kept its brightness at 
a constant level, showing just pronounced wave-like 
orbitally-related variations 
\citep[][ this paper]{sk+07,sk+12} -- the main characteristic 
of the quiescent phase of a symbiotic star. 

During June 2015, AG~Peg experienced a new outburst 
(AAVSO Alert Notice 521
\footnote{https://www.aavso.org/aavso-alert-notice-521}). 
On July 1, 2015, \cite{steele+15} indicated strong emission 
lines of, namely, \ion{H}{i}, \ion{He}{ii,} and \ion{N}{iii} 
superposed on nebular continuum with a pronounced Balmer jump 
in emission on their low-resolution spectrum (320--620\,nm). 
Around a maximum of the optical brightness, from June 28 to 
July 8, 2015, the Swift/XRT observation detected an increase 
in the count rate with a factor of $\sim$3 with respect to the 
previous observation on August 16--19, 2013 
\citep[][]{nunez+13,luna+15}. 
Following Swift/XRT, monitoring of the 2015 outburst indicated 
a markedly variable X-ray flux on a time scale of days 
\citep[][]{ramsay+15,ramsay+16}. The authors ascribed 
the origin of the strong and variable X-rays to shocks in 
the variable ejecta. 
Analysing the same Swift/XRT observations, \cite{zh+tom16} 
came to the conclusion that the characteristics of the X-ray 
emission from the 2015 outburst could not result from colliding 
stellar winds in a binary system as was indicated by previous 
\textsl{ROSAT} observations from June of 1993. 
Using optical spectroscopy and $BV$ photometry, \cite{tomov+16} 
concluded that the 2015 outburst of AG~Peg places it among the 
classical Z~And-type symbiotic binary systems. 

In this contribution we determine basic characteristics and 
the nature of the 2015 AG~Peg outburst using a high cadence 
optical spectroscopy and multicolour $UBVR_{\rm C}I_{\rm C}$ 
photometry complemented with ultraviolet spectra from 
the preceding quiescence. 
In Sect.~\ref{sec:obs} we summarise and describe our observations 
and data reduction. Section~\ref{sec:analysis} describes our 
analysis and presents the results. Their discussion and summary 
are found in Sects.~\ref{sec:dis} and \ref{sec:sum}, respectively. 

\section{Observations}
\label{sec:obs}
\subsection{Photometry}
Multicolour photometric observations were carried out with 
two 60cm, f/12.5 Cassegrain telescopes at the Star\'{a} 
Lesn\'{a} Observatory (G1 and G2 observing pavilions) operated 
by the Astronomical Institute of the Slovak Academy of Sciences. 

1. CCD photometry (G1 pavilion) was obtained in the standard 
Johnson-Cousins $UBVR_{\rm C}I_{\rm C}$ system using the 
\textsl{FLI ML3041} CCD camera (2048$\times$2048 px, 
pixel size: 15\,$\mu$m $\times $15\,$\mu$m, scale: 0.4 arcsec/px, 
FoV: $14^{\prime}\times 14^{\prime}$), 
mounted at the Cassegrain focus. The data were reduced using 
the \textsl{IRAF} software package\footnote{http://iraf.noao.edu} 
as described by \cite{pv05}. All frames were dark-subtracted, 
flat-fielded and corrected for cosmic rays. 
Corresponding magnitudes were obtained using the comparison 
stars "a", "b", and "c" of \cite{henden+06}. 
To get a better coverage, we complemented our CCD measurements 
with those from the AAVSO database 
\footnote{https://www.aavso.org/data-download}. 

2. Photoelectric $UBV$ photometry was carried out in the G2 
pavilion by a single-channel photometer mounted in the Cassegrain 
focus \citep[see][ in detail]{vanko+15a,vanko+15b}. 
The star HD~207\,933 ($V$ = 8.10, $B-V$ = 1.05, $U-B$ = 0.97) 
was used as the comparison star. 
Internal uncertainties of these one-day-mean measurements are 
of a maximum of a few $\times 0.01$\,mag in $U$, but smaller in 
other filters. Our photometric measurements are summarised in 
Table~\ref{tab:phot}. 

Multicolour photometry was used mainly to convert the relative 
and/or arbitrary flux units of the low-resolution and/or the 
medium-resolution spectra to absolute fluxes. 
To obtain flux-points of the true continuum, we determined 
corrections for emission lines by using our low-resolution 
spectra \citep[see][]{sk07}. Magnitudes were converted to fluxes 
according to the calibration of \cite{hk82} and \cite{bessel79}. 

\subsection{Spectroscopy}
\label{sec:spec}
Spectroscopic observations were secured at different
observatories and/or private stations: 

(i)
At the Star\'{a} Lesn\'{a} Observatory in the G1 observing 
pavilion with a 0.6\,m Cassegrain telescope equipped with an 
eShel fiber-fed spectrograph (Shelyak; optical fiber of 50\,$\mu$m) 
mounted at the f/5 focus. 
The detector was an ATIK\,460EX CCD camera using $2750\times 2200$ 
chip with 4.54\,$\mu$m square pixels and $2\times 2$ binning. 
Reduction of the spectroscopy was performed using dedicated 
scripts written using the \textsl{IRAF} package tasks, Linux shell 
scripts, and \textsl{FORTRAN} programs 
\citep[see][]{pribulla+15}. 

(ii)
At the Ond\v{r}ejov Observatory, medium-resolution 
spectroscopy was performed using the coude single dispersion  
slit spectrograph of 2\,m reflector and the {\small BROR} CCD     
camera with the SITe-005 800$\times$2030\,pixels chip. 
The resolution power at the \ha\ region was 13\,000. 
Standard initial reduction (bias subtraction, flat-fielding and 
wavelength calibration) was carried out using modified 
\textsl{IRAF} packages (MW). 

(iii)
At the Crimean Astrophysical Observatory with the 2.6\,m 
Shajn telescope, using a \textsl{SPEM} slit spectrograph mounted 
at the Nasmith focus. The detector was a SPEC-10 CCD camera 
with the 1340$\times$100\,pixels chip. Average resolution was 
R = 1000. 
The primary reduction of the spectra, including the bias subtraction 
and flat fielding, was performed with the \textsl{SPERED} code 
developed by S.I. Sergeev at the Crimean Astrophysical Observatory 
(TNT). 
 
(iv)
At the private station in Rouen with a 0.36\,m Schmidt-Cassegrain 
telescope (Meade) equipped with an eShel spectrograph (Shelyak, 
optical fiber of 50\,$\mu$m) mounted at the f/d5.2 focus using 
a reducer. The detector was an ATIK\,460EX CCD camera (pixel 
size of 4.54\,$\mu$m, binning 2$\times$2 mode; pixel of 
9.08\,$\mu$m) (FT). 

(v)
At the Fujii Kurosaki Observatory with a 0.4\,m 
Schmidt-Cassegrain telescope F10 (Meade), using a 
{\small FBSPEC-III} spectrograph and a CCD camera ML6303E(FLI) 
(3072$\times$2048 pixel) as the detector (MF). 

(vi) 
At the Santa Maria de Montmagastrell Observatory, T\`arrega 
(Lleida) Spain, using a Control Remote Telescope SC16 equipped 
with a spectrograph {\small B60050-VI}. The detector was an 
ATIK\,460EX CCD camera (JG). 

(vii)
At the Observatory de la Tourbi\`ere with a 0.4\,m 
Ritchey-Chr\'etien telescope (Astrosib RC400) equipped with 
an eShel spectrograph (Shelyak, optical fiber of 50\,$\mu$m) 
mounted at the f/d5.5 focus. The detector was an ATIK\,460EX 
CCD camera (pixel size of 4.54\,$\mu$m, binning 2$\times$2 
mode; pixel of 9.08\,$\mu$m) (OG). 

(viii)
At the private stations, Manhattan (Illinois) and Grand 
Lake (Colorado), by a 0.25\,m LX200 Schmidt-Cassegrain telescope 
using a Shelyak {\small Alpy 600} spectrograph. The detector 
was an ATIK\,314L CCD camera (KG). 

(ix)
At the private station Mill Ridge with a 0.31\,m Cassegrain 
telescope equipped with a classical slit spectrograph coupled with 
an f/d6.6 reducer to the f/d10 focus. The spectrograph was used 
with a 23\,$\mu$m slit (2.3 arcsec) and a 1800 l/mm grating in 
the first order. The detector was a QSI583 (KAF8300 chip) binned 
1$\times$2 giving 0.33\,\AA/pixel (R = 9000 at \ha) (T. Lester). 

(x)
At the Observatoire de Haute-Provence with a 0.2\,m Newtonian 
telescope (Orion N200 f/5) equipped with a Shelyak 
{\small Alpy 600} spectrograph. The detector was an 
ATIK\,460EX CCD camera (Sony ICX694 sensor, binning 1$\times$1) 
(VB). 

(xi) 
At the private station Chelles, with a 0.35\,m SCT F11 (Celestron) 
telescope, using the eShell cross-dispersed echelle spectrograph 
(Shelyak) and a CCD camera ATIK\,460EX with ICX694 (Sony) sensor 
(T. Lemoult). 

(xii)
At the private station in L'Aquila with a 0.20\,m 
Schmidt-Cassegrain telescope f/10, a custom built spectrograph, 
equipped with a 600 l/mm grism. The detector was 
a SBIG ST-8300M CCD camera (US). 

(xiii)
At the Observatory de la Couy\`ere (CALC UAI code J23) 
with a 0.355\,m Schmidt-Cassegrain telescope 
(Meade ACF $14^{\prime\prime}$ f/d10) equipped with an 
{\small Alpy 600} Shelyak spectrograph with 23\,$\mu$m slit. 
The detector was an ATIK\,460EX CCD camera (pixel size of 
4.54\,$\mu$m, binning 1$\times$1) (JM). 

(xiv)
At the West Challow Observatory using a 0.28\,m Schmidt-Cassegrain 
telescope (Celestron C11) at f/5 with a LISA spectrograph (Shelyak) 
and SXVR-H694 CCD camera (DB). 

Our optical spectroscopy was complemented with the 
UV/optical and far-UV spectroscopy. The former was carried 
out by the Faint Object Spectrograph (\textsl{FOS}) onboard 
the {\em Hubble Space Telescope} (\textsl{HST}) during 
November 11, 1993 (spectra: Y1JO0305T, Y1JO0308T, Y1JO0309T, 
Y1JO0406T) and during December 1996 (spectra: Y3CK530AT, 
Y3KK040FT, Y3KK040BT, Y3KK0406T, Y3KK0408T). 
The latter was carried out by the {\em Far Ultraviolet 
Spectroscopic Explorer} (\textsl{FUSE}) on June 5.618, 2003 
(the spectrum q1110103). We used the calibrated time-tag 
spectrum taken through MDRS aperture in the LiF1A 
(988--1082\,\AA) channel. 

Low-resolution spectra ($R\sim 500-1000$, Table~\ref{tab:low}), 
preferentially those covering the Balmer discontinuity, were 
used for modelling the SED. Absolute flux calibration and 
correctness of the spectra were verified with the aid of the 
(near-)simultaneous $UBVR_{\rm C}I_{\rm C}$ photometry and with 
the reduced $\chi^2$ quantity of the resulting model. 
Medium-resolution spectra ($R\sim 11000$, Table~\ref{tab:med}) 
served to analyse variations in the line profiles. Absolute 
fluxes of the used parts of the spectrum were obtained by linear 
interpolation between models of the two nearest low-resolution 
spectra. 
The \textsl{HST/FOS} spectra were used for modelling the SED 
of the UV/optical continuum, whereas the \textsl{FUSE} spectrum 
was used to analyse the \ion{O}{vi} 1032\,\AA\ line. 
Observations were dereddened with $E_{\rm B-V}$ = 0.1 using 
the extinction curve of \cite{c+89} and resulting parameters 
were scaled to a distance of 800\,pc 
\citep[see][ and Appendix~A in detail]{kenyon+93}. 
Finally, we determine the orbital phase of the binary according 
to the ephemeris of the inferior conjunction of the cool giant 
as \citep[see][]{fekel+00}, 
\begin{equation}
 JD_{\rm spec. conj.} = 
    2\,447\,165.3(\pm 48) + 818.2(\pm 1.6)\times E .
\end{equation}
%
%
%
%
\begin{table}
\caption[]{Photometric observations of AG~Peg. The full table 
           is available in electronic form via CDS.}
\begin{center}
\begin{tabular}{ccccccr}
\hline
\hline
JD~2\,4... &   $U$ &   $B$ &  $V$  &$R_{\rm C}$&$I_{\rm C}$& Note$^{1}$\\
\hline
 56997.22&  9.775&  9.994&  8.884&     --&    -- &  PP \\
\multicolumn{7}{c}{..............................................} \\
 57281.29&  7.809&  8.667&  7.926&  6.819&  6.087& CCD \\
 57395.21&  7.769&  8.713&  8.000&     --&    -- &  PP \\
\hline
\end{tabular}
\end{center}
  $^{1}$~A charge-coupled device (CCD) or 
         photoelectric (PP) photometry
\label{tab:phot}
\end{table}
%
%
\begin{table}
\caption[]{Log of low-resolution spectroscopic observations}
\begin{center}
\begin{tabular}{@{~~}c@{~~~}c@{~~~}c@{~~~}r@{~~~}r@{~~~}c@{~~}}
\hline
\hline
Date$^{a}$ & JD~2\,4...  & Region&$T_{\rm exp}$& R$^{b}$&Observatory$^{c}$\\
yyyy/mm/dd.ddd &         &  [nm] &   [s]       &        &        \\
\hline
2013/10/27.087 & 56592.587 & 363-740 &4267 & 500 & (viii) \\
2015/06/26.955 & 57200.455 & 370-737 &1936 & 500 & (xii)  \\
2015/07/16.926 & 57220.426 & 330-757 &  40 &1000 & (iii)  \\
2015/07/22.977 & 57226.477 & 330-757 &  40 &1000 & (iii)  \\
2015/07/23.937 & 57227.437 & 330-757 &  20 &1000 & (iii)  \\
2015/07/24.973 & 57228.473 & 330-757 &  60 &1000 & (iii)  \\
2015/08/02.014 & 57236.514 & 378-747 & 917 & 984 & (vi)   \\
2015/08/17.983 & 57252.483 & 354-831 & 417 & 695 & (x)    \\
2015/08/19.946 & 57254.446 & 333-755 &  90 &1000 & (iii)  \\
2015/08/20.933 & 57255.433 & 331-758 &  30 &1000 & (iii)  \\
2015/08/21.890 & 57256.390 & 329-757 &  90 &1000 & (iii)  \\
2015/09/04.538 & 57270.038 & 355-860 & 500 & 500 & (v)    \\
2015/09/19.846 & 57285.346 & 330-758 &  30 &1000 & (iii)  \\
2015/09/26.950 & 57292.450 & 370-740 &1322 & 637 & (xiii) \\
2015/10/11.539 & 57307.039 & 355-860 & 500 & 500 & (v)    \\
2015/10/12.589 & 57308.089 & 355-860 & 750 & 500 & (v)    \\
2015/10/13.442 & 57308.942 & 355-860 & 600 & 500 & (v)    \\
2015/10/19.804 & 57315.304 & 330-757 &  20 &1000 & (iii)  \\
2015/10/21.465 & 57316.965 & 355-860 & 600 & 500 & (v)    \\
2015/10/29.724 & 57325.224 & 330-757 &  30 &1000 & (iii)  \\
2015/10/30.682 & 57326.182 & 330-757 &  90 &1000 & (iii)  \\
2015/11/08.037 & 57334.537 & 373-738 & 265 & 525 & (viii) \\
2015/11/15.453 & 57341.953 & 360-860 & 350 & 500 & (v)    \\
2015/11/23.882 & 57350.382 & 378-747 & 344 & 976 & (vi)   \\
2015/11/29.790 & 57356.290 & 378-747 & 282 & 925 & (vi)   \\
2015/12/08.453 & 57364.953 & 360-860 & 600 & 500 & (v)    \\
2015/12/24.790 & 57381.290 & 390-740 &1012 & 821 & (xiv)  \\
2016/01/02.761 & 57390.261 & 385-753 & 239 & 772 & (vi)   \\
\hline
\end{tabular}
\end{center}
{\bf Notes.}
  $^{(a)}$~Start of the observation in UT,
  $^{(b)}$~average resolution, 
  $^{c}$~according to the list in Sect.~\ref{sec:spec}. 
\label{tab:low}
\end{table}
%
%
%
\begin{table}
\caption[]{Log of medium resolution spectroscopic
           observations$^{a}$}
\begin{center}
\begin{tabular}{cccrc}
\hline
\hline
Date$^{b}$  & JD~2\,4... & Region & $T_{\rm exp}$ & Observatory$^{d}$  \\
yyyy/mm/dd.ddd &           &   [nm] &   [s] &           \\
\hline
2006/10/18.926 & 54027.426 & 641-692 & 2000 & (ii) \\
2013/08/11.900 & 56516.400 & 427-761 & 1885 & (xi) \\
2013/08/29.945 & 56534.445 & 641-688 & 1401 & (ii)\\
2013/09/04.843 & 56540.343 & 641-688 &  301 & (ii) \\
2013/09/04.855 & 56540.355 & 466-489 &  801 & (ii) \\
2013/09/27.978 & 56563.478 & 466-489 &  901 & (ii) \\
2013/10/07.004 & 56572.504 & 641-688 &  901 & (ii) \\
2013/10/24.874 & 56590.374 & 466-489 &  601 & (ii) \\
2013/10/26.793 & 56592.293 & 466-489 & 2001 & (ii) \\
2013/12/17.716 & 56644.216 & 466-489 & 3601 & (ii) \\
2015/06/27.227 & 57200.727 & 601-711 & 2047 & (ix)  \\
2015/07/01.011 & 57204.511 & 418-731 & 2747 & (vii) \\ 
2015/07/09.999 & 57213.499 & 422-710 & 3600$^{c}$ & (i)  \\
2015/07/17.004 & 57220.504 & 422-710 & 3600$^{c}$ & (i) \\
2015/07/29.942 & 57233.442 & 421-715 & 1106 & (iv) \\
2015/08/04.971 & 57239.471 & 422-710 & 3600$^{c}$ & (i) \\
2015/08/11.006 & 57245.506 & 641-688 &  801$^{c}$ & (ii) \\
2015/08/13.896 & 57248.396 & 641-688 &  601 & (ii) \\ 
2015/08/21.896 & 57256.396 & 421-715 & 1262 & (iv) \\
2015/08/21.953 & 57256.453 & 641-688 &  421 & (ii) \\
2015/08/23.002 & 57257.502 & 422-710 &  720 & (i) \\
2015/08/23.983 & 57258.483 & 422-710 &  900 & (i) \\
2015/08/25.039 & 57259.539 & 422-710 &  600 & (i) \\
2015/08/28.927 & 57263.427 & 421-715 & 1798 & (iv) \\
2015/09/06.883 & 57272.383 & 421-715 &  684 & (iv) \\
2015/09/10.852 & 57276.352 & 421-715 &  710 & (iv) \\
2015/09/19.875 & 57285.375 & 421-715 &  692 & (iv) \\
2015/09/25.837 & 57291.337 & 421-715 &  445 & (iv) \\
2015/10/01.825 & 57297.325 & 421-715 & 1661 & (iv) \\
2015/10/08.849 & 57304.349 & 421-715 &  985 & (iv) \\
2015/10/11.779 & 57307.279 & 421-715 &  888 & (iv) \\
2015/10/12.131 & 57307.631 & 600-710 & 1175 & (ix)  \\
2015/10/13.777 & 57309.277 & 421-715 & 1197 & (iv) \\
2015/10/20.830 & 57316.330 & 421-715 &  678 & (iv) \\
2015/10/25.747 & 57321.247 & 421-715 & 1408 & (iv) \\
2015/10/27.780 & 57323.280 & 422-710 & 1440 & (i) \\
2015/11/01.771 & 57328.271 & 421-715 &  579 & (iv) \\
2015/11/01.807 & 57328.307 & 418-731 & 5533 & (vii) \\
2015/11/01.817 & 57328.317 & 641-688 & 2001 & (ii) \\
2015/11/01.840 & 57328.340 & 466-489 & 1207 & (ii) \\
2015/11/13.749 & 57340.249 & 421-715 & 1271 & (iv) \\
2015/11/25.801 & 57352.301 & 421-715 &  836 & (iv) \\
2015/12/07.805 & 57364.305 & 421-715 &  679 & (iv) \\
2015/12/12.803 & 57369.303 & 421-715 & 1214 & (iv) \\
2015/12/23.735 & 57380.235 & 421-715 &  520 & (iv) \\
2015/12/29.761 & 57386.261 & 421-715 &  702 & (iv) \\
2016/01/08.720 & 57396.220 & 422-710 & 1800 & (i) \\
2016/01/13.723 & 57401.223 & 421-715 & 1458 & (iv) \\
\hline
\end{tabular}
\end{center}
{\bf Notes.}
  $^{a}$~ Average resolution is 11000, 
  $^{b}$~ start of the observation in UT, 
  $^{c}$~ \ha\ saturated, 
  $^{d}$~according to the list in Sect.~\ref{sec:spec}. 
\label{tab:med}
\end{table}
%
%
%
\begin{table}
\caption[]{Parameters from model SEDs: ST, $T_{\rm e}$ (K), 
           $EM$ (10$^{60}$\cmt) and a minimum of the 
           $\chi_{\rm red}^2$ function 
           (see Sect.~\ref{sec:sed}).}
\begin{center}
\begin{tabular}{clccc}
\hline
\hline
Date           &  ST  & $T_{\rm e}$ & $EM$ & $\chi_{\rm red}^2$/d.o.f. \\
\hline
\multicolumn{5}{c}{Quiescent phase} \\
\hline
1993/11/13.463$^{a}$ & 3.5$^{d}$ & 15000 & 0.329 & --         \\
1996/12/12.209$^{b}$ & 3.5       & 28000 & 0.188 & 0.630/1648 \\
2013/10/27.087       & 2.6       & 17500 & 0.740 & 1.265/1396 \\
\hline
\multicolumn{5}{c}{Active phase} \\ 
\hline
2015/06/26.955 & 1.4 & 30000$^{c}$ & 9.11  & 0.649/1006\\
2015/07/16.926 & 1.4 & 23500       & 5.93  & 1.410/1689\\
2015/07/22.977 & 2.2 & 22000       & 4.69  & 1.193/1618\\
2015/07/23.937 & 1.5 & 26500       & 4.72  & 0.739/1468\\
2015/07/24.973 & 2.3 & 22000       & 4.57  & 1.118/1667\\
2015/08/02.014 & 1.7 & 26000$^{c}$ & 3.91  & 0.841/3720\\
2015/08/17.983 & 1.8 & 21000       & 1.89  & 1.183/6091\\
2015/08/19.946 & 1.8 & 19500       & 2.42  & 1.518/1536\\
2015/08/20.933 & 2.5 & 19000       & 2.35  & 1.163/1544\\
2015/08/21.890 & 2.3 & 19500       & 2.36  & 1.693/1590\\
2015/09/04.538 & 2.1 & 19500$^{c}$ & 2.14  & 1.073/901\\ 
2015/09/19.846 & 2.5 & 20000       & 2.31  & 1.616/1685\\
2015/09/26.950 & 1.7 & 22000$^{c}$ & 2.32  & 0.790/4835\\
2015/10/11.539 & 1.9 & 24500       & 5.59  & 0.763/1062\\
2015/10/12.589 & 2.1 & 26000       & 5.89  & 0.881/1019\\
2015/10/13.442 & 2.1 & 26000       & 5.97  & 1.000/1113\\
2015/10/19.804 & 1.6 & 26500       & 4.42  & 0.806/1671\\
2015/10/21.465 & 1.6 & 26500       & 6.27  & 1.214/1053\\
2015/10/29.724 & 1.8 & 23500       & 4.78  & 0.938/1672\\
2015/10/30.682 & 2.3 & 23000       & 4.56  & 0.930/1684\\
2015/11/08.037 & 2.3$^{d}$ & 23000$^{c}$ & 4.33 & 0.930/2142\\
2015/11/15.453 & 1.9 & 24000       & 4.06  & 0.804/979\\
2015/11/23.882 & 2.3$^{d}$& 23000$^{c}$ & 3.90  & 1.449/1312\\
2015/11/29.790 & 2.3$^{d}$& 23000$^{c}$ & 3.19  & 1.020/3716\\
2015/12/08.453 & 2.3$^{d}$& 15000       & 2.30  & 0.734/1109\\
2015/12/24.790 & 2.3$^{d}$& 23000$^{c}$ & 1.82  & 0.755/2640\\
2016/01/02.761 & 2.3$^{d}$& 19000$^{c}$ & 1.53  & 1.275/4322\\
\hline
\end{tabular}
\end{center}
{\bf Notes.} 
$^{(a)}$~ Hot component: 
          $T_{\rm h} \equiv 1.5\times 10^5$\,K, 
          $R_{\rm h} = 0.088$\ro, $L_{\rm h} = 3560$\lo; 
$^{(b)}$~ $T_{\rm h} \equiv 1.5\times 10^5$\,K,         
          $R_{\rm h} = 0.071$\ro, $L_{\rm h} = 2280$\lo. 
$^{(c)}$~ $T_{\rm e}$ estimated using the photometric flux in $U$; 
$^{(d)}$~ adopted value. 
\label{tab:sed}
\end{table}  

\section{Analysis and results}
\label{sec:analysis}

\subsection{Photometric evolution}
\label{sec:phot}

\begin{figure}
\begin{center}
\resizebox{\hsize}{!}{\includegraphics[angle=-90]{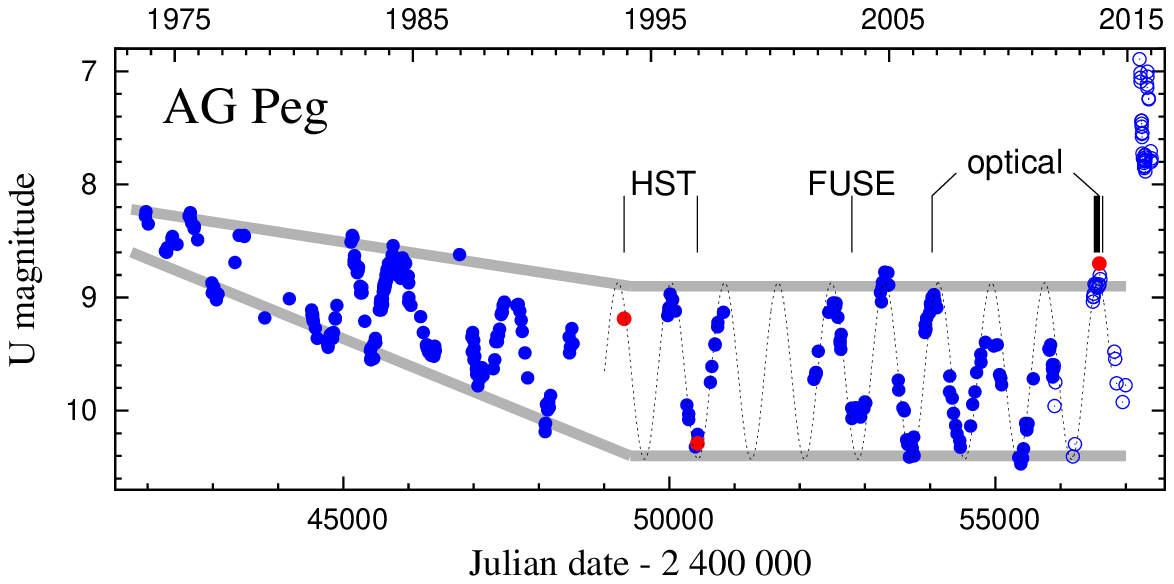}}
\end{center}
\caption[]{The final stage of the nova-like outburst of AG~Peg 
as demonstrated by the $U$-\textsl{LC} (blue dots) from 1972 
to the present. Vertical lines represent timing of our 
spectroscopic observations during the quiescent phase. 
Red dots represent $U$ magnitudes derived from model SEDs 
(see Sect.~\ref{sec:em}). 
Data are from \cite{belyakina92}, \cite{mn94}, \cite{hric+96}, 
\cite{tt98}, \cite{sk+04}, \cite{sk+07}, \cite{sk+12} and 
this paper ($\odot$). 
          }
\label{fig:ulc}
\end{figure}

Figure~\ref{fig:ulc} shows the $U$-\textsl{LC} of AG~Peg, which 
demonstrates that its nova-like outburst terminated around 
1997, when the overall decline ended and the orbitally 
related wave-like variation reached its maximum 
($\Delta U \sim 1.5$\,mag), varying around 9.7\,mag until June 2015. 
The pronounced wave like variability in the optical continuum 
represents a distinctive photometric feature of quiescent phases 
of symbiotic binaries \citep[e.g.][]{hoff68,puc70,mein79}. 
This periodic variation is given by apparent changes of the 
nebular component of radiation as a function of the orbital 
phase \citep[see Fig.~3 of][ and Sect.~\ref{sec:em} below]{sk08}. 
According to the STB model, these changes can be caused by 
the densest parts of the ionised wind from the giant, which is 
partially optically thick and located in between the binary 
components \citep[see][]{sk01}. 
During outbursts stages, the major source of the nebular 
radiation is represented by the ionised wind from the hot 
component, whose emissivity is not orbitally related, but 
follows the mass-loss rate \citep[see][]{sk06}. From this 
point of view, the increasing amplitude of the wave-like 
variation along the decline of AG~Peg is a result of weakening 
of the nebular emission from the hot star's wind, which is 
consistent with the observed decrease of the mass-loss rate 
from the hot component (Sect.~\ref{sec:intr}). 
Transition of AG~Peg to quiescent phase at some point during 
1995--2000 (see Fig.~\ref{fig:ulc}) thus triggered 
the accretion process -- a fundamental condition for the new 
energetic event. 
%

During June of 2015, AG~Peg commenced a new active phase 
(see Figs.~\ref{fig:ulc} and \ref{fig:ubvri}). 
According to $B$ and $V$ CCD measurements from the AAVSO 
database and those collected by the VSOLJ observer Hiroyuki 
Maehara, the brightening started on June 5.10$\pm 3.3$, 2015 
(JD~2\,457\,177.6$\pm 3.3$) at $B = 9.55\pm 0.07$ and 
$V = 8.45\pm 0.06$, and reached a maximum on June 30.0$\pm 2.0$, 
2015 (JD~2\,457\,203.5$\pm 2.0$) at $B = 7.68\pm 0.05$ and 
$V = 7.0\pm 0.1$\,mag. 
A three months gradual decline was interrupted by a weaker 
secondary brightening by $\Delta U\approx 1$, $\Delta B\sim 0.7$ 
and $\Delta V\sim 0.5$\,mag on October 8.5$\pm 0.5$, 2015 
(JD~2\,457\,304.0$\pm 0.5$) on the time-scale of 1 day. Then, 
the \textsl{LC}s showed a plateau phase until November 23, 2015, 
after which the brightness gradually decreased to values similar 
to those prior to the secondary eruption, at the end of the observing 
season of AG~Peg in January 2016 (see Fig.~\ref{fig:ubvri}). 
Such evolution in the multicolour \textsl{LC} of symbiotic binaries 
(brightening, timescale, and multiplicity) is classified as 
the Z~And-type outburst. 
%
%

\subsection{Spectroscopic evolution}
\label{sec:param}

In this section we determine physical parameters by modelling 
the continuum and analysing the line spectrum. We describe 
their temporal evolution along the outburst. Evolution of 
the main spectroscopic and physical parameters is depicted 
in Figs.~\ref{fig:param} and \ref{fig:lrt}. Corresponding data 
are included in Tables~\ref{tab:sed}, \ref{tab:lrt}, and 6. 

\subsubsection{Modelling the SED} 
\label{sec:sed}

According to properties of symbiotic stars 
(see Sect.~\ref{sec:intr}), their observed spectrum, $F(\lambda)$, 
can be expressed as a superposition of its three basic 
components, 
%
\begin{equation}
  F(\lambda) =  F_{\rm h}(\lambda) + F_{\rm n}(\lambda) +
                F_{\rm g}(\lambda),
\label{eq:sed1}  
\end{equation}
where $F_{\rm h}(\lambda)$, $F_{\rm n}(\lambda),$ and 
$F_{\rm g}(\lambda)$ represent radiative contributions from 
the hot component, nebula, and giant, respectively. Because of 
the very high temperature of the hot component 
($T_{\rm h}\sim 10^5$\,K, Sect.~\ref{sec:intr}), we observe only 
its long-wavelength tail in the UV/optical range, which can be 
approximated by a blackbody radiation. 
The nebular continuum was ascribed to processes of recombination 
and thermal bremsstrahlung in the hydrogen plasma for Case~B, 
because no feature of the He$^{+}$ and He$^{+2}$ 
nebular continuum was recognisable in the spectrum 
(e.g. the 2051\,\AA\ discontinuity) and all the continua are 
similar in the profile \citep[][]{bm70}. 
Finally, the radiation from the giant was compared with a 
synthetic spectrum, $\mathcal{F}_{\lambda}(T_{\rm eff})$, 
according to models of \cite{fluks+94}. 
Then Eq.~(\ref{eq:sed1}) can be expressed as \citep[][]{sk05b}, 
%
\begin{equation}
 F(\lambda) =   
      \theta_{\rm h}^2 \pi B_{\lambda}(T_{\rm h}) + 
      k_{\rm n} \varepsilon_{\lambda}(T_{\rm e}) + 
      \mathcal{F}_{\lambda}(T_{\rm eff}), 
\label{eq:sed2}
\end{equation}
%
where the angular radius of the hot component, 
$\theta_{\rm h} = R_{\rm h}/d$, 
is given by its effective radius, $R_{\rm h}$ 
(i.e. the radius of a sphere with the same luminosity) 
and the distance $d$. 
The factor $k_{\rm n}$ [cm$^{-5}$] scales the volume emission
coefficient $\varepsilon_{\lambda}(T_{\rm e})$ 
[${\rm erg\,cm^3\,s^{-1}\,\AA^{-1}}$] of the nebular continuum
to observations. Further, the electron temperature, $T_{\rm e}$, 
and thus $\varepsilon_{\lambda}(T_{\rm e})$ are assumed to be 
constant throughout the nebula, which simplifies determination 
of the emission measure of the nebula to 
\textsl{EM} = $4\pi d^2 k_{\rm n}$\cmt\ 
\citep[see Eq.~(9) of][]{sk15}. 
Angular radius of the giant, $\theta_{\rm g}$, can be 
obtained from its observed bolometric flux, 
%
\begin{equation}
F_{\rm g}^{\rm obs} = \int_{\lambda}\!
         \mathcal{F}_{\lambda}(T_{\rm eff})\,{\rm d}\lambda  =
         \theta_{\rm g}^2\!\!\int_{\lambda}\!
         \pi B_{\lambda}(T_{\rm eff})\,{\rm d}\lambda  = 
         \theta_{\rm g}^2 \sigma T_{\rm eff}^{4} .
\label{eq:fbol}
\end{equation}
The variables determining the model SED are $\theta_{\rm h}$, 
$T_{\rm h}$, $k_{\rm n}$, $T_{\rm e}$, $\theta_{\rm g}$ and 
$T_{\rm eff}$. 
In the SED-fitting analysis, we compared a grid of models 
(\ref{eq:sed2}) with the observed continuum, and selected 
the one corresponding to a minimum of the reduced $\chi^2$ 
function. More details can be found in \cite{sk05b,sk15}. 

In modelling the UV/optical continuum from 1993 and 1996, we 
adopted $T_{\rm h} = 150000$\,K as indicated during quiescent 
phase (see Sect.~\ref{sec:th}). 
For the 1996 spectrum, we simultaneously fitted 133 continuum 
fluxes from 1144 to 3644\,\AA\  with 1520 flux-points from 
3743 to 5635\,\AA. In this way we obtained parameters, 
$\theta_{\rm h}$, $k_{\rm n}$, $T_{\rm e}$ and the spectral type 
ST (i.e. $T_{\rm eff}$) of the giant according to calibration 
of \cite{fluks+94}. For the 1993 spectrum, we fitted only 15 
continuum fluxes between 1280 and 3600\,\AA, because of very rich 
emission-line spectrum creating a bump between 2200 and 2600\,\AA. 
Here, parameters $\theta_{\rm h}$, $k_{\rm n}$ and $T_{\rm e}$ 
were obtained. To match also a short optical part of the spectrum 
($\lambda < 4700$\,\AA), the synthetic spectrum from the 1996 model 
was scaled to the observed continuum by eye. Therefore, we do 
not infer 
the value of $\chi_{\rm red}^2$ for this model SED. 

To fit just the optical continuum, the method is able to 
disentangle contributions only from the nebula and the RG, 
because the WD radiation is negligibly small in the optical at 
the observed very high temperatures. Therefore, it was possible 
to determine only parameters, $k_{\rm n}$, $T_{\rm e}$, and ST. 
We also determined a subclass of the ST by a linear interpolation 
between the neighbouring best-fitting STs. Because the radiation 
from the giant can vary, the scaling factor of the synthetic 
spectra represents another parameter in the model SED. 

The continuum fluxes were selected by eye. Within the 
optical, we distinguished the line-free regions with the aid 
of the synthetic spectrum of the giant. Spectral regions 
between the Balmer jump and $\sim$3800\,\AA, from $\sim$4720 
to $\sim$4830\,\AA,\ and from $\sim$6860 to $\sim$6960\,\AA\ 
were omitted from modelling because of blends of hydrogen 
emission lines, a large difference from the synthetic spectrum, 
and the water vapour absorptions in the Earth's atmosphere, 
respectively. 
A close comparison of the observed and synthetic spectrum 
revealed uncertainties of the spectral calibration to 5--10\% 
between $\sim$4200 and $\sim$7100\,\AA, while the ends of most 
spectra were inclined from their model prediction up to 20\% 
(see Fig.~\ref{fig:sedopt}). 
Considering spectral ranges with a better and worse calibration, 
and for the sake of simplicity, we adopted uncertainties 
of selected continuum fluxes at all low-resolution spectra 
of 7\%. 

The resulting parameters are listed in Table~\ref{tab:sed} and 
examples of model SEDs are depicted in Figs.~\ref{fig:seduv} and 
\ref{fig:sedopt}. 
%
%
\begin{figure}
\begin{center}
\resizebox{\hsize}{!}{\includegraphics[angle=-90]{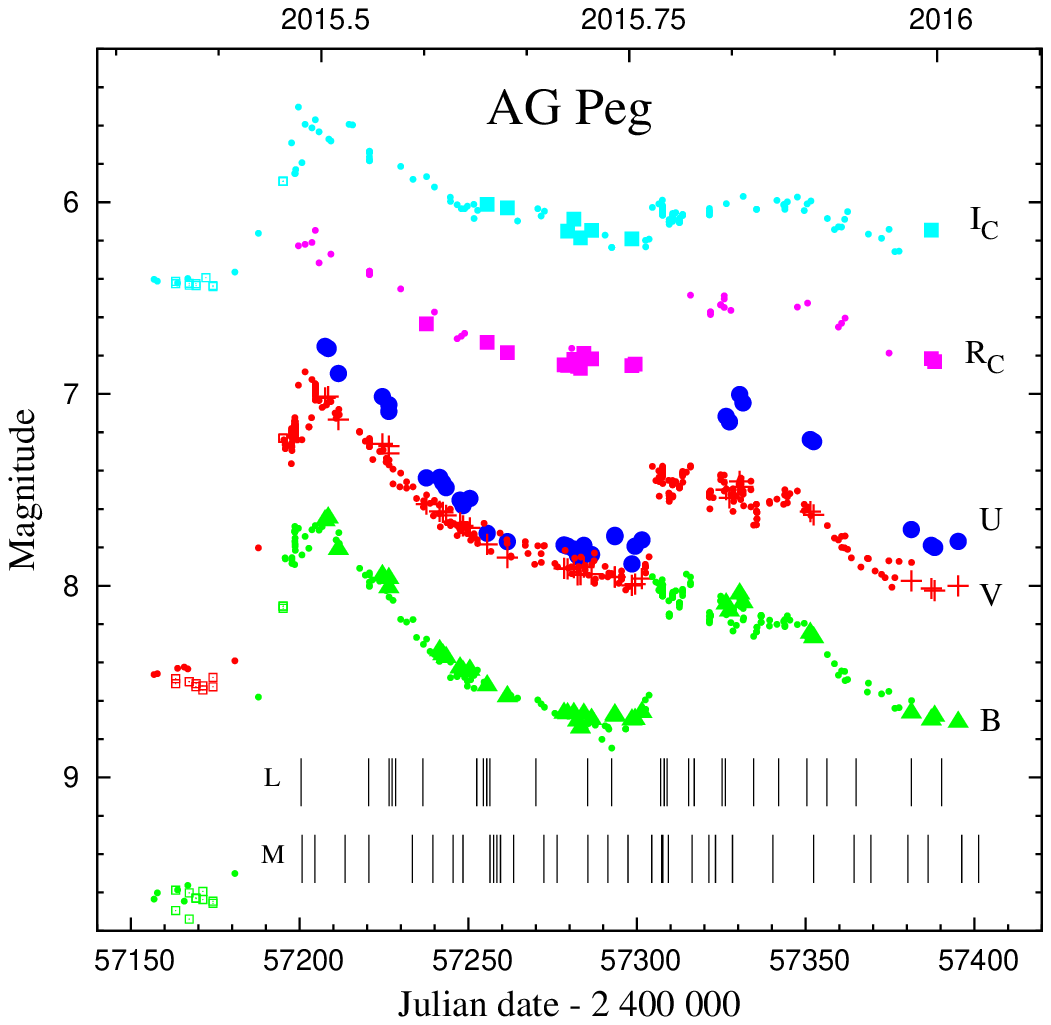}}
\end{center}
\caption[]{$U,B,V,R_{\rm C},I_{\rm C}$ \textsl{LC}s of AG~Peg 
during its 2015 outburst. Large symbols are our measurements 
(Table~\ref{tab:phot}), while small ones are from the AAVSO 
database and those collected by the VSOLJ observer Hiroyuki
Maehara (open squares prior to the brightening). 
Vertical bars denote the dates of our low (L) and medium (M) 
resolution spectra. 
          }
\label{fig:ubvri}
\end{figure}
%
%
%
\begin{figure}
\begin{center}
\resizebox{\hsize}{!}{\includegraphics[angle=-90]{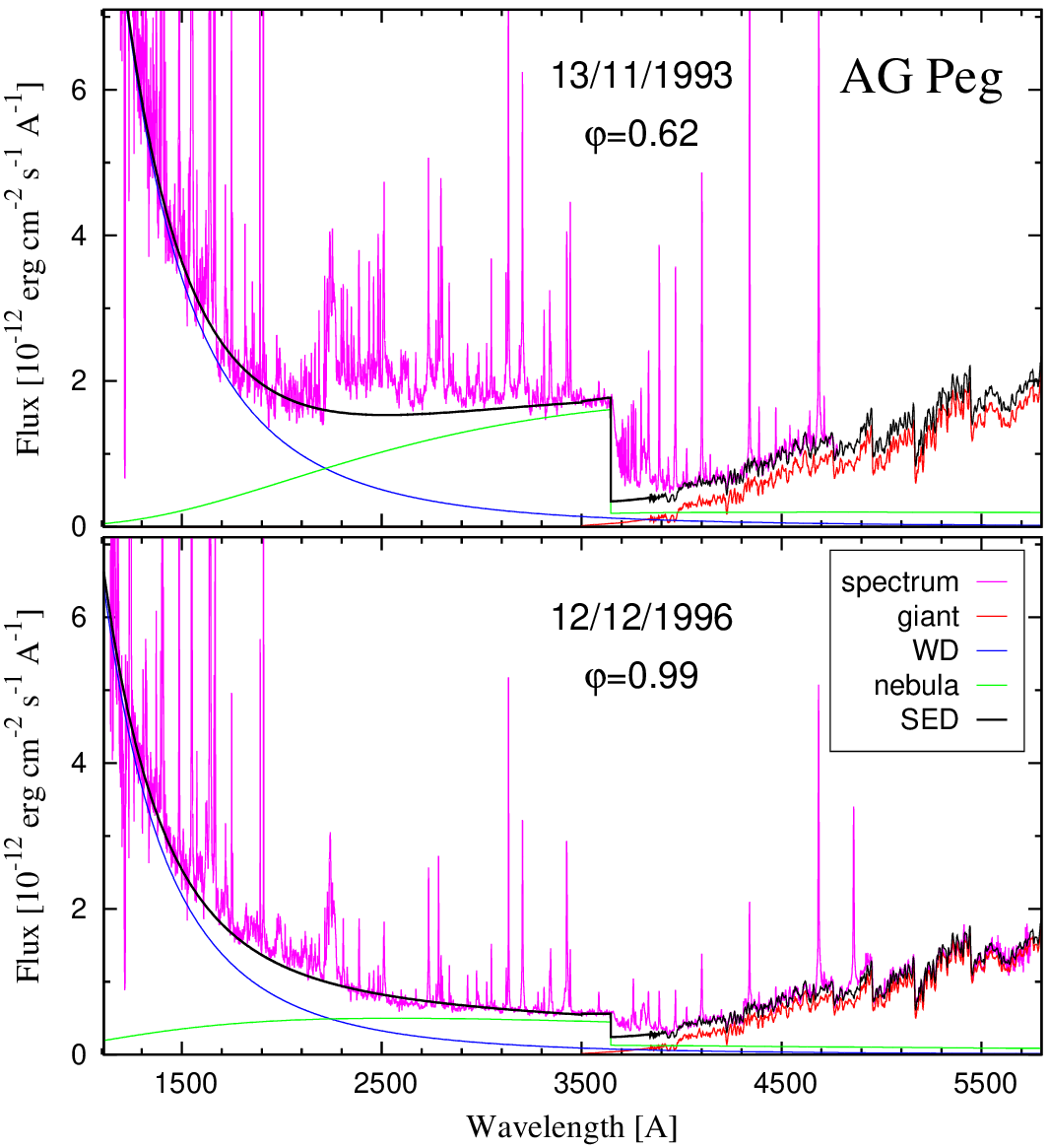}}
\end{center}
\caption[]{
UV/optical SED of AG~Peg during quiescent phase from two 
opposite orbital phases -- around the inferior and superior 
conjunction of the giant. This demonstrates the origin 
of the orbitally-related light variation in the $U$ band 
shown in Fig.~\ref{fig:ulc}. 
          }
\label{fig:seduv}
\end{figure}

\subsubsection{Emission measure}
\label{sec:em}

\begin{figure*}
\begin{center}
\resizebox{18cm}{!}{\includegraphics[angle=-90]{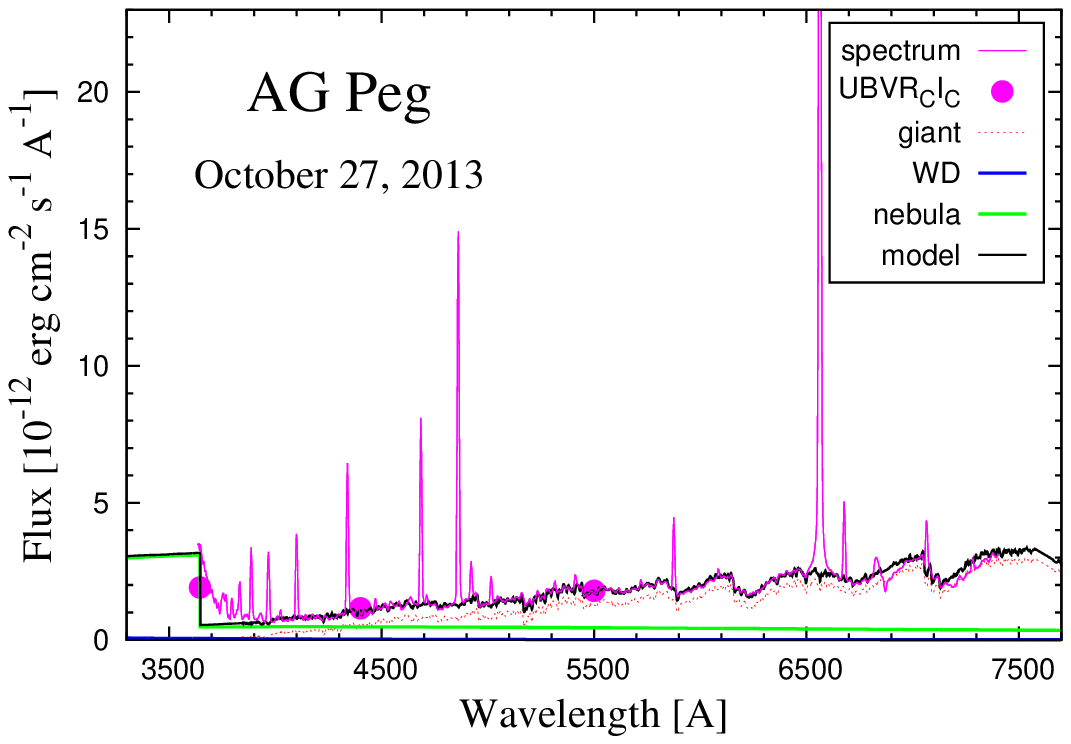}
                    \includegraphics[angle=-90]{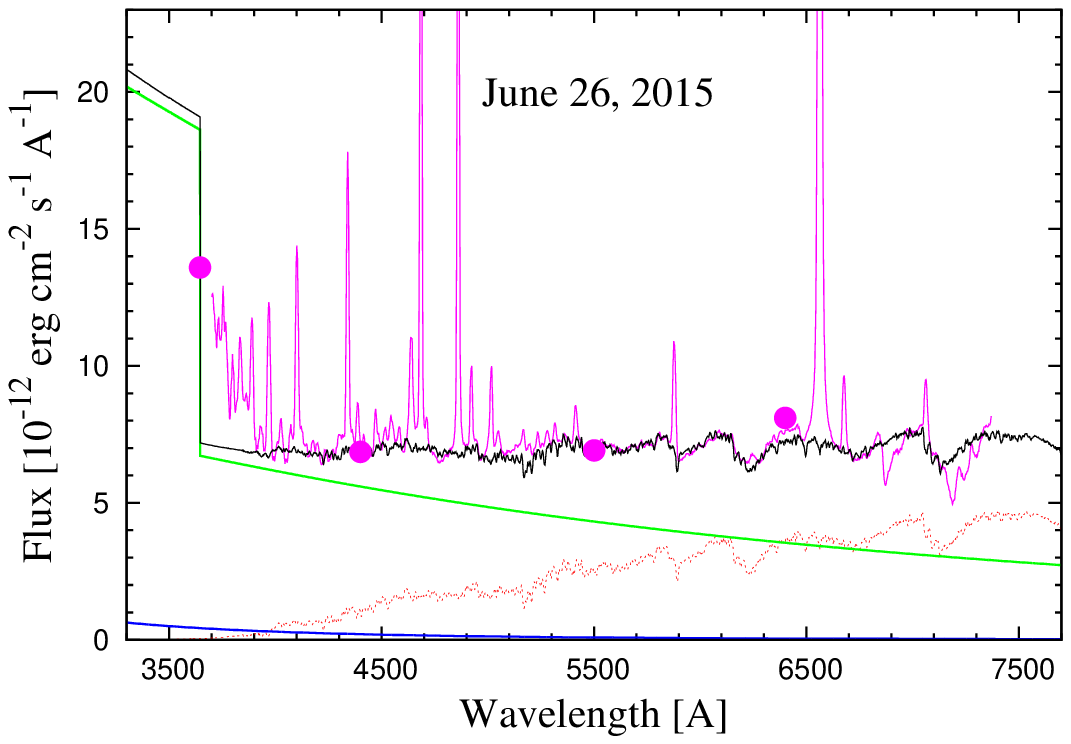}
                    \includegraphics[angle=-90]{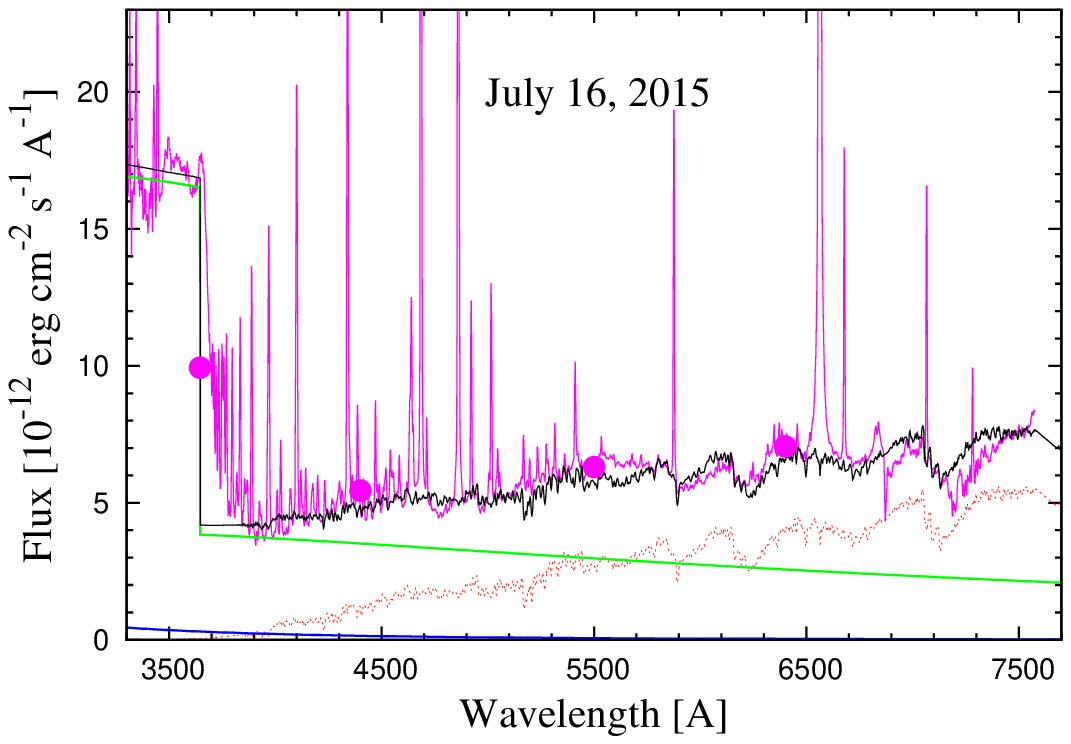}}
\resizebox{18cm}{!}{\includegraphics[angle=-90]{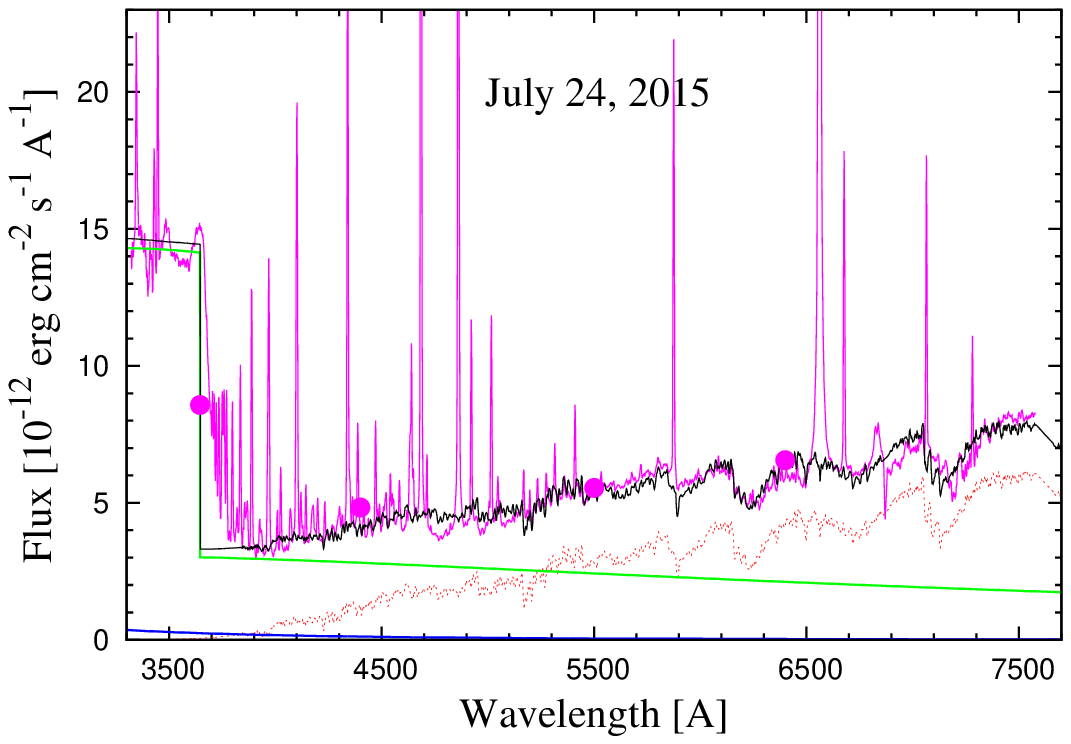}
                    \includegraphics[angle=-90]{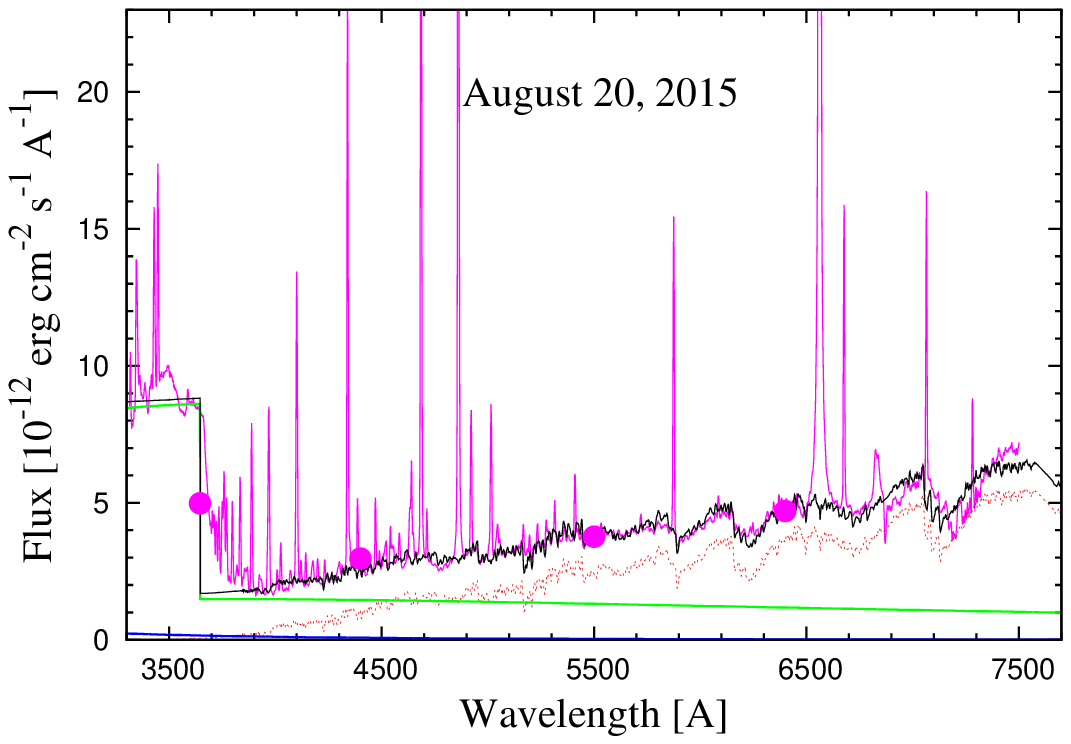}
                    \includegraphics[angle=-90]{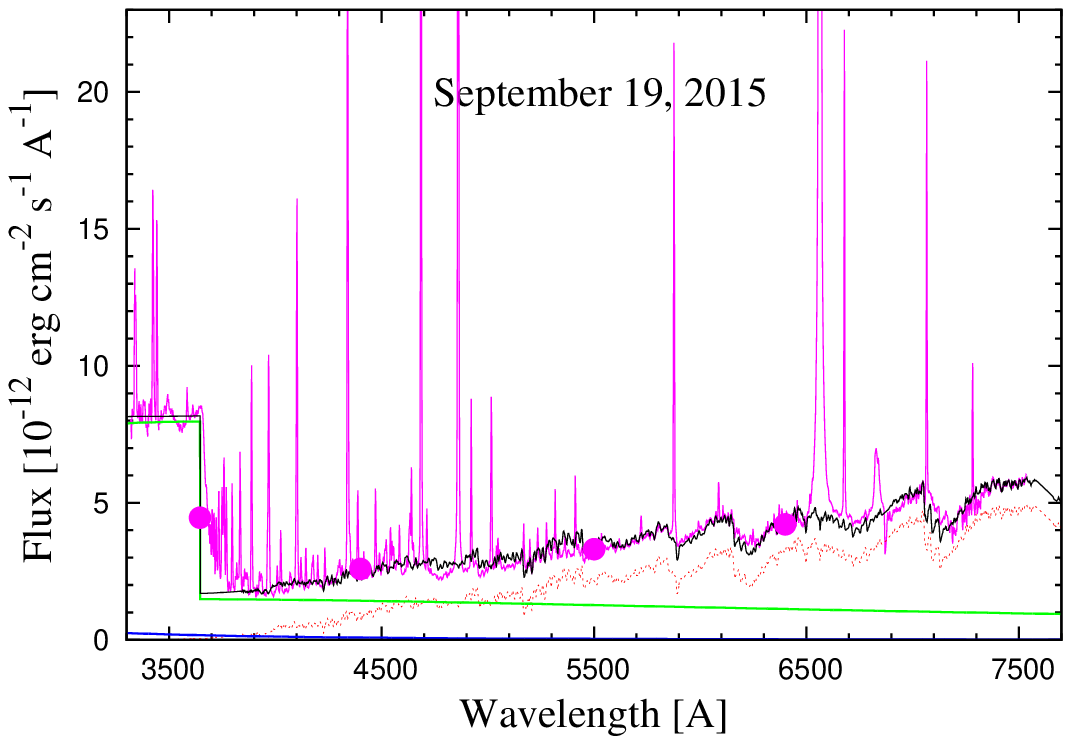}}
\resizebox{18cm}{!}{\includegraphics[angle=-90]{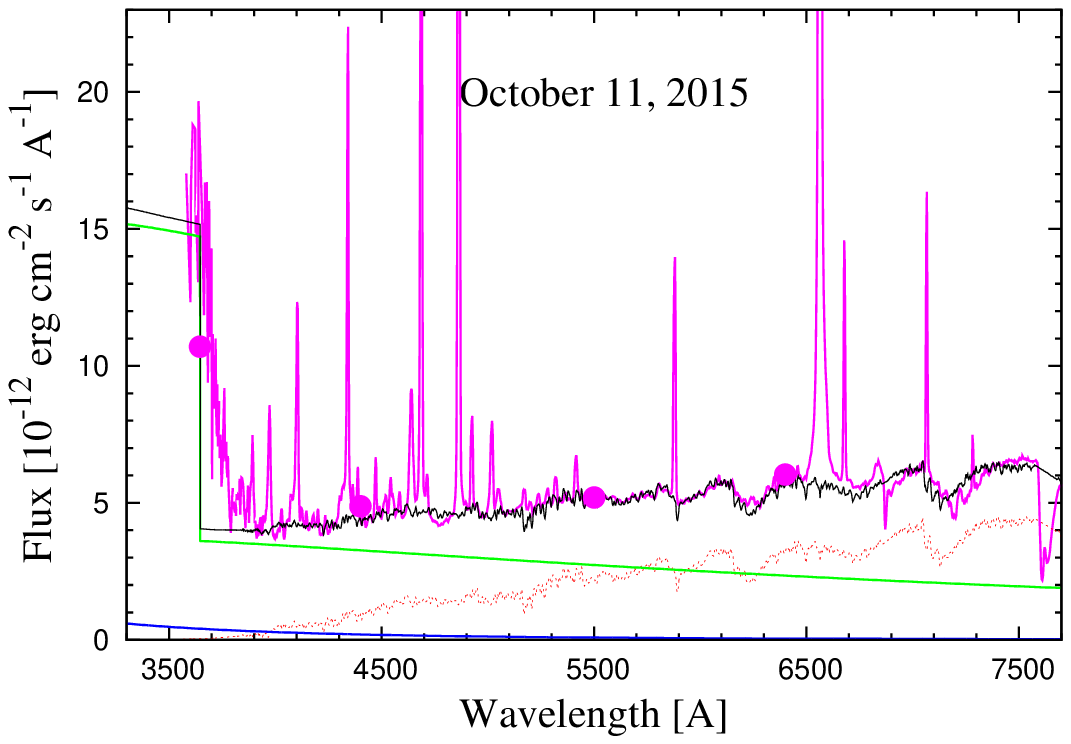}
                    \includegraphics[angle=-90]{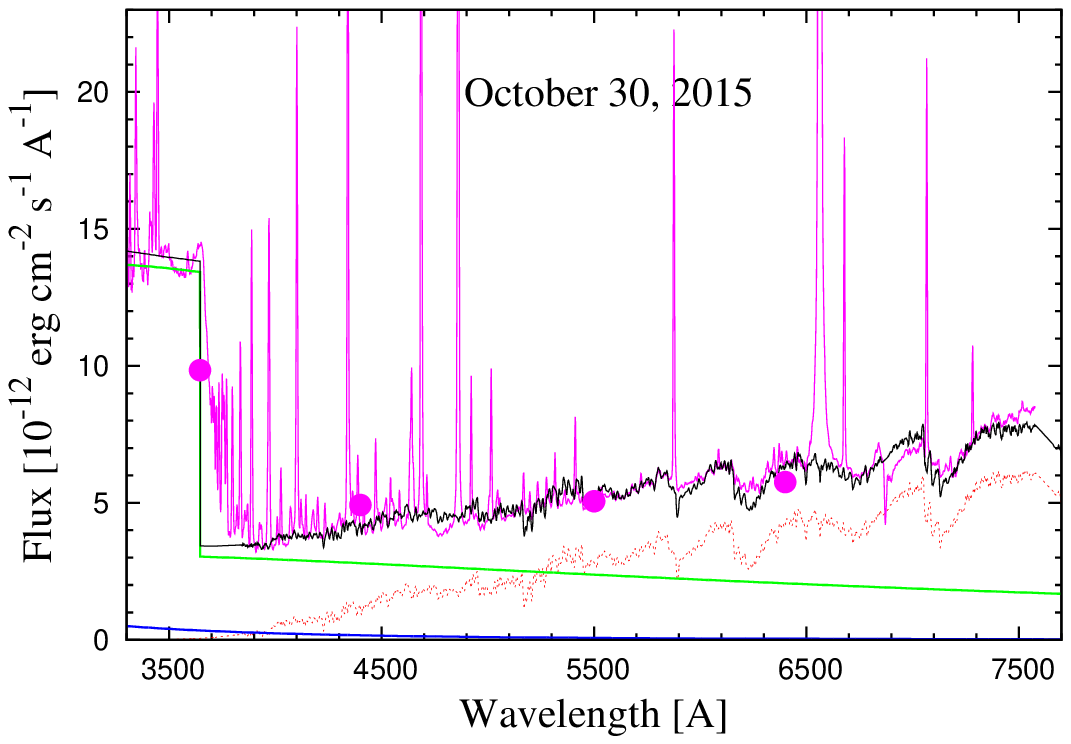}
                    \includegraphics[angle=-90]{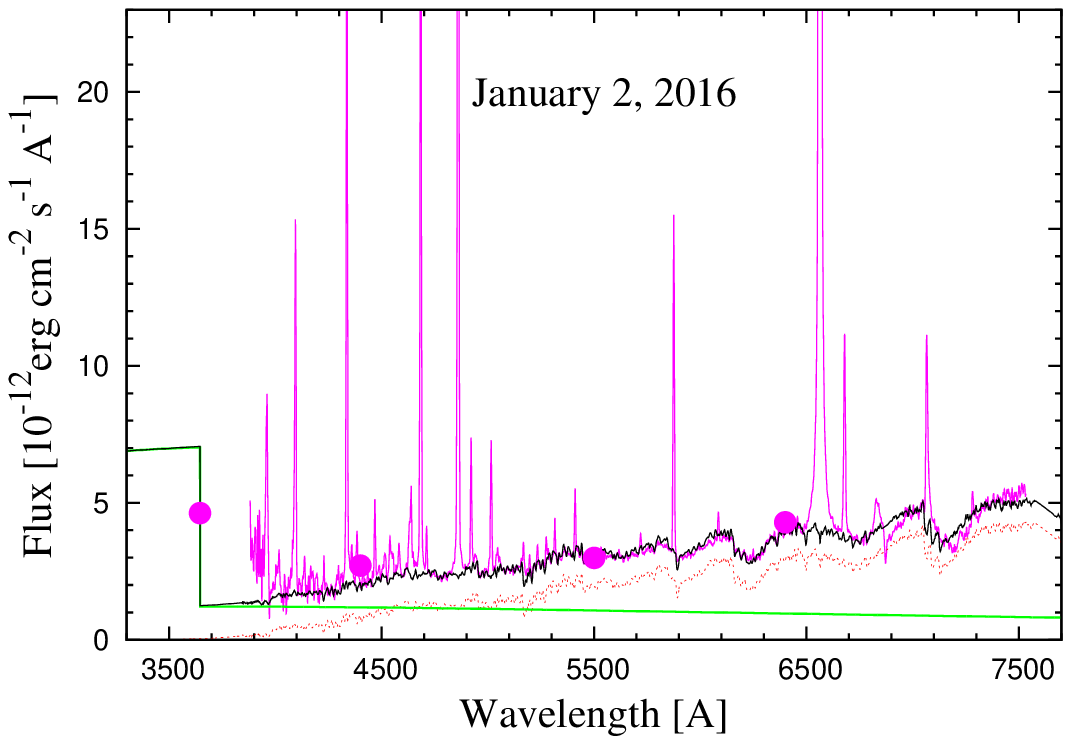}}
\end{center}
\caption[]{
Examples of our low-resolution dereddened spectra (magenta lines) 
and their model SEDs (black lines) at selected dates prior to and 
during the 2015 outburst of AG~Peg. Denotation of lines is as in 
Fig.~\ref{fig:seduv}. 
Contribution from the WD corresponds to parameters in 
Table~\ref{tab:lrt}. It is negligibly small throughout 
the optical. Modelling is described in Sect.~\ref{sec:sed}. 
          }
\label{fig:sedopt}
\end{figure*}

Our model SEDs revealed the presence of a strong nebular 
component of radiation during both the quiescent and active 
phase of AG~Peg. It dominated the spectrum from the near-UV 
to around the $B$ passband 
(see Figs.~\ref{fig:seduv} and \ref{fig:sedopt}). 

During quiescent phase, $EM$ varies with the orbital phase. 
This is demonstrated by model SEDs of the \textsl{HST} spectra 
taken at the orbital phase $\varphi = 0.62$ and 0.99 
(Fig.~\ref{fig:seduv}, Table~\ref{tab:sed}). If we convert the 
model fluxes at the $U$ band to the scale of the observed $U$ 
magnitudes, we get an excellent agreement with the photometric 
$U$-\textsl{LC} (see the red points in Fig.~\ref{fig:ulc}). 
To get the proper quantity of the nebular contribution, we 
considered the emission coefficient $\varepsilon_{U}$ as the 
weighted average of its values from both sides of the Balmer 
jump, 
$\varepsilon_{U}=0.6\varepsilon_{U^{-}}+0.4\varepsilon_{U^{+}}$ 
\citep[see][]{cs10}, multiplied by the model parameter 
$k_{\rm n}$ (Eq.~(\ref{eq:sed2})). 
Adding model contributions from the giant and hot component, we 
obtained magnitudes corresponding to the true dereddened continuum. 
To get the observed magnitudes, we added a correction for the 
emission lines, $\Delta U_{\it l} = 0.19$\,mag \citep[][]{sk07} 
and the reddening of +0.48\,mag in the $U$ passband. In this way 
we obtained $U$ = 9.18 and 10.29 for the above mentioned orbital 
phases. Similarly, we obtained $U \sim 8.7$ from the model SED 
on October 27, 2013 ($\varphi = 0.52$). 
This result confirms that the periodic wave-like variation 
in the $U$-\textsl{LC} of AG~Peg from $\sim 1997$ was caused by the 
apparent orbital changes in the $EM$, similarly to variation 
during quiescent phases of other symbiotic stars 
\citep[see][ and Sect.~\ref{sec:phot} above]{sk01}. 

During the outburst, $EM$ increased by a factor of $\sim 10$, 
and its quantity followed the \textsl{LC} profile 
(see Fig.~\ref{fig:param}). 
No orbitally related variation could be recognised. This implies 
that the large nebular emission was produced by a different source 
than during quiescence. According to the ionisation structure of 
hot components during active phases \citep[][]{cs12}, and 
a significant broadening of the \ion{H}{i} and \ion{He}{ii} line 
profiles (Fig.~\ref{fig:broad}), the large $EM$ was generated by 
the ionised wind from the hot component \citep[][]{sk06}. 
%
%

\subsubsection{\ion{He}{ii} 4686\,\AA\ and \hb\ lines}
\label{sec:he2h}
\begin{figure}
\begin{center}
\resizebox{\hsize}{!}{\includegraphics[angle=-90]{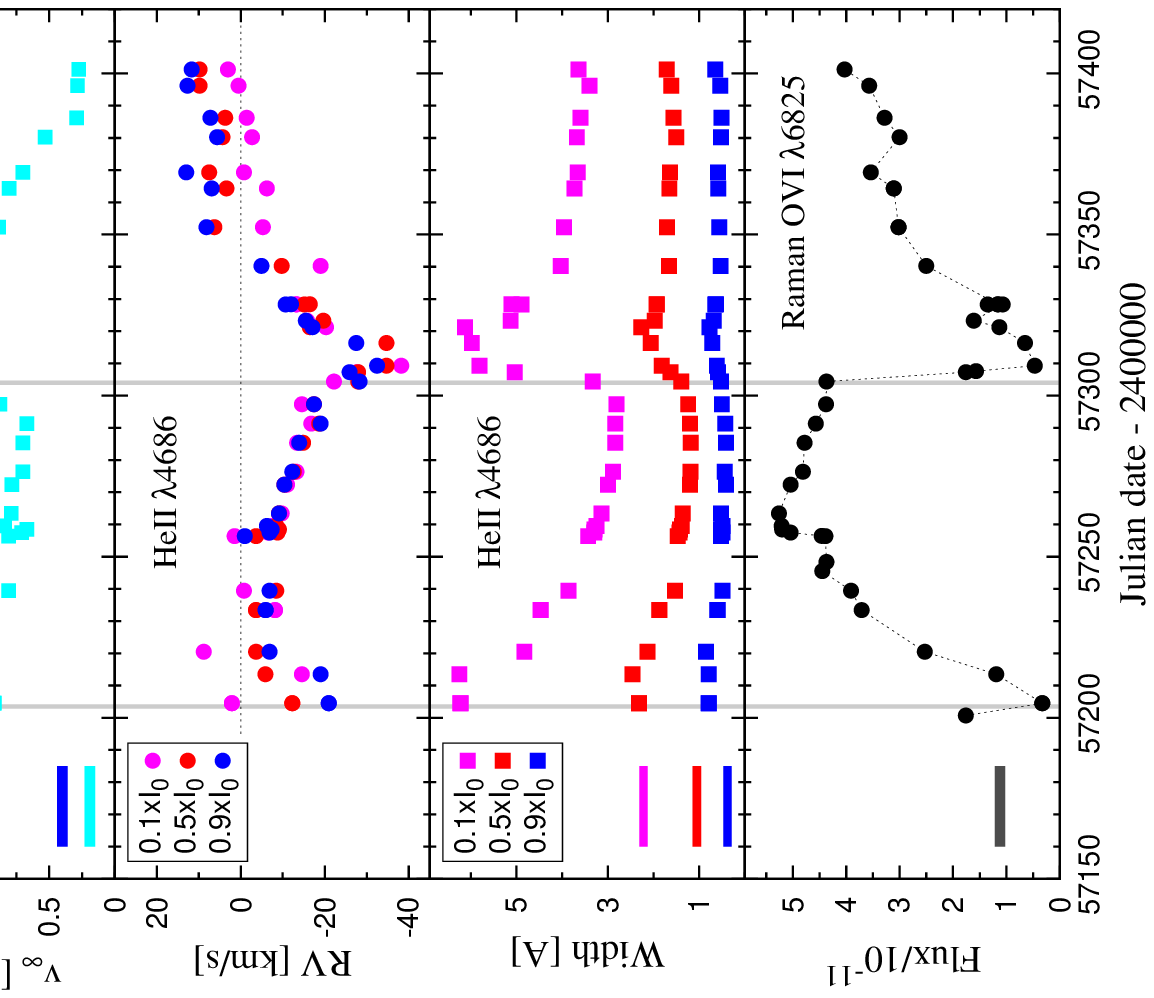}}
\end{center}
\caption[]{
Evolution of \textsl{EM} (full squares -- Table~\ref{tab:sed}; 
open squares -- interpolated to dates of medium-resolution 
spectra) and parameters of \ion{He}{ii}\,$\lambda 4686$, \hb, 
\ha\ and Raman-scattered \ion{O}{vi}\,$\lambda 6825$ lines 
along the 2015 outburst of AG~Peg. 
The horizontal belts at the bottom left corners denote the average 
quantity of the parameter during the preceding quiescence. 
Fluxes (in \ecs) are listed in Table~6. 
          }
\label{fig:param}
\end{figure}

Basic properties of the symbiotic nebula during the outburst 
-- the ionised wind from the hot component -- can be probed 
by analysing recombination lines, \ion{He}{ii} 4686\,\AA\ and 
\hb. The zones of their ions can be different in size, and 
subject to variation during the outburst. Both lines were well 
exposed at all our spectra. 
The \ion{He}{ii}\,$\lambda$4686 profile was more or less 
symmetrical, whereas the \hb\ emission was attenuated by 
an absorption component from its blue side. Therefore, 
to estimate the total emission of the \hb\ line, we 
removed the absorption component by fitting the emission core 
with Gaussian curves. 
Evolution of \ion{He}{ii}\,$\lambda$4686 and \hb\ parameters 
is depicted in Fig.~\ref{fig:param}, and can be summarised 
as follows. 

\begin{figure}
\begin{center}
\resizebox{\hsize}{!}{\includegraphics[angle=-90]{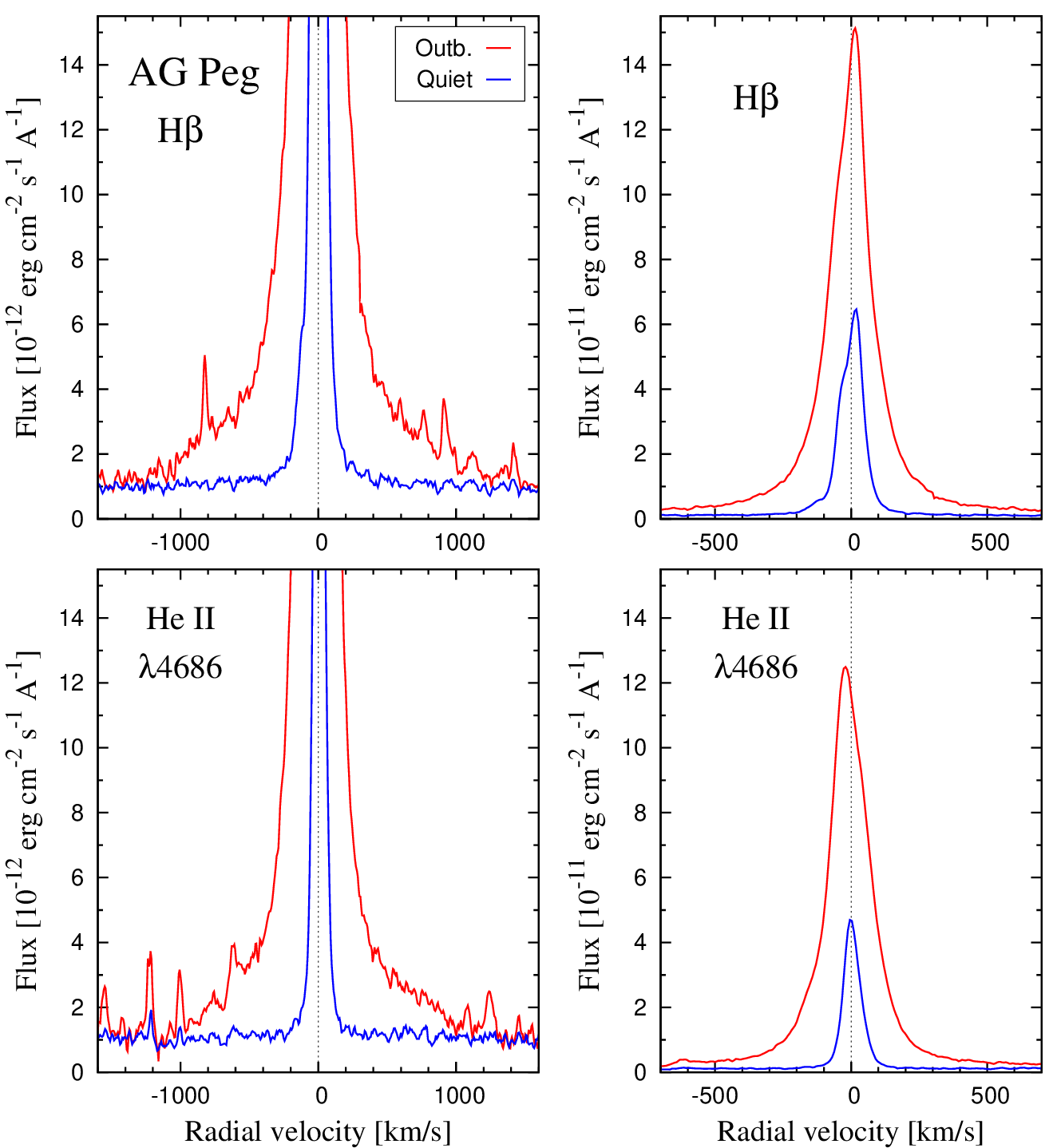}}
\end{center}
\caption[]{Significant broadening of \hb\ and 
\ion{He}{ii}\,$\lambda$4686 line profiles was observed during 
the outburst (red line, July 9, 2015) with respect to quiescence 
(blue line, October 26, 2013). Local continuum was subtracted. 
          }
\label{fig:broad}
\end{figure}

(i) 
During the outburst, fluxes of \ion{He}{ii} 4686\,\AA\ and \hb\ 
lines increased by a factor of $\ga 5$ with respect to values 
from the preceding quiescence. Their variation basically copied 
that in \textsl{LC}s and \textsl{EM}, but with 
a less steep increase during the June/July and October 2015 
brightening. A significant broadening of their profiles immediately 
after the optical maximum is depicted in Fig.~\ref{fig:broad}. 

(ii)
A distinctive change was observed in their broad wings. During 
quiescence, no broad wings in the \ion{He}{ii} 4686\,\AA\ 
profile were clearly seen. We measured just the core emission 
with HWZI$\sim 200$\kms. In the \hb\ profile, weak broad wings 
terminated at $v_{\infty}\approx \pm 400$\kms. 
During the outburst, broad wings expanding to the terminal 
velocity $v_{\infty}\ga 1000$\kms\ developed in both the lines. 
In the \hb\ profile, $v_{\infty}$ did not change significantly 
for the whole observing period, whereas $v_{\infty}$ of 
\ion{He}{ii} 4686\,\AA\ was variable and, at the end of 2015, 
decreased to values of the quiescent phase. This suggests 
a variable size of the He$^{+2}$ zone during the outburst 
(see Sect.~\ref{dis:he2} in detail). 

(iii)
Broadening of the whole \ion{He}{ii} 4686\,\AA\ profile (we 
measured its width at 0.1, 0.5 and 0.9$\times I_{\rm max}$) 
was significant at/after the optical maxima. For example, the 
width at 0.1$\times I_{\rm max}$ increased with a factor of 
$\sim 3$ relative to quantities of the quiescent phase. 
The broadening of the line correlates with the evolution in
the \textsl{EM} and $\dot{M}_{\rm h}$ (Figs.~\ref{fig:param}
and \ref{fig:lrt}), which justifies that the nebular emission 
was produced by the enhanced ionised wind from the WD during
the outburst. 

(iv)
The position of the profile shifted temporarily towards negative 
radial velocities (RVs) with a maximum of $\sim -35$\kms\ just 
after the October brightening, and returned back to the position 
of its reference wavelength at the end of November 2015 
(third panel from the bottom). 
This interesting effect is probably connected with the ionisation 
structure of the hot component and its evolution in the course of 
the outburst (see Sect.~\ref{dis:shift} in detail). 
%
%
%
%
%

\subsubsection{Temperature of the ionising source}
\label{sec:th}

Based on the fact that the presence of ions $X^{\rm +i}$ in the 
nebula requires the presence of photons with 
$h\nu > \chi_{\rm i}$, where $\chi_{\rm i}$ is the ionisation 
energy necessary to ionise $X^{\rm +(i-1)}$ to $X^{\rm +i}$, 
\cite{mn94} found an empirical relation between the temperature 
of the ionising source, $T_{\rm h}$, and the ionisation 
potential $\chi_{\rm i}$ of the ion $X^{\rm +i}$ as 
\begin{equation}
  \frac{T_{\rm h}}{\chi_{\rm i}} = 1000 ~ {\rm [K/eV]}. 
\label{eq:tis}
\end{equation}
Thus the presence of an ion with the highest ionisation energy, 
$\chi_{\rm max}$, determines the lower limit of $T_{\rm h}$. 
Using this approach to the spectrum of AG~Peg, the presence 
of the \ion{O}{vi} 1032\,\AA\ line and/or its Raman scattered 
counterpart at 6825\,\AA\ (see Sect.~\ref{sec:ovi}) requires 
$T_{\rm h} \ga 114\,000$\,K, because 
$\chi_{\rm max}$ = $\chi$(O$^{+5}) \sim 114$\,eV. 

Having only optical spectra available, we can use the 
\ion{He}{ii}($\lambda 4686$)/\hb\ flux ratio to estimate 
$T_{\rm h}$. Under optically thick conditions, where 
the $F_{4686}$ nebular flux growths from complete absorption 
of stellar photons solely by He$^{+}$ ions, 
$Q(4\nu_0,\infty)$, and the $F_{\rm H\beta}$ nebular flux results 
from complete absorption of photons between the frequency $\nu_0$ 
and $4\nu_0$, $Q(\nu_0,4\nu_0)$, a relationship for the flux 
ratio can be expressed as, 
\begin{equation}
 \frac{Q(4\nu_0,\infty)}{Q(\nu_0,4\nu_0)} = 
 \frac{\alpha({\rm H}\beta)}{\alpha_{4686}}
 \frac{\alpha_{\rm B}({\rm He}^{+})}{\alpha_{\rm B}({\rm H}^0)}
 \frac{\nu_{4861}}{\nu_{4686}}
 \frac{F_{4686}}{F_{{\rm H}\beta}}, 
\label{eq:heh}
\end{equation}
where $\nu_0$ is the ionising frequency of hydrogen 
\citep[e.g.][]{gurz97}. $\alpha({\rm H}\beta)$ and $\alpha_{4686}$ 
are effective recombination coefficients for the given transition, 
whereas $\alpha_{\rm B}({\rm He}^{+})$ and 
$\alpha_{\rm B}({\rm H}^0)$ are total recombination coefficients 
of He$^{+}$ and H$^0$ for the Case B. Dominant 
absorption of photons by He$^{+}$ ions within the innermost 
He$^{+2}$ zone is due to recombinations of He$^{+2}$ 
to He$^{+}$ that produce sufficient quanta to keep the hydrogen 
ionised \citep[][]{hs64}. This procedure was first suggested by 
\cite{ambartsumyan} and later modified by more authors 
\citep[e.g.][]{iijima,kj89}. 

\begin{figure}
\begin{center}
\resizebox{\hsize}{!}{\includegraphics[angle=-90]{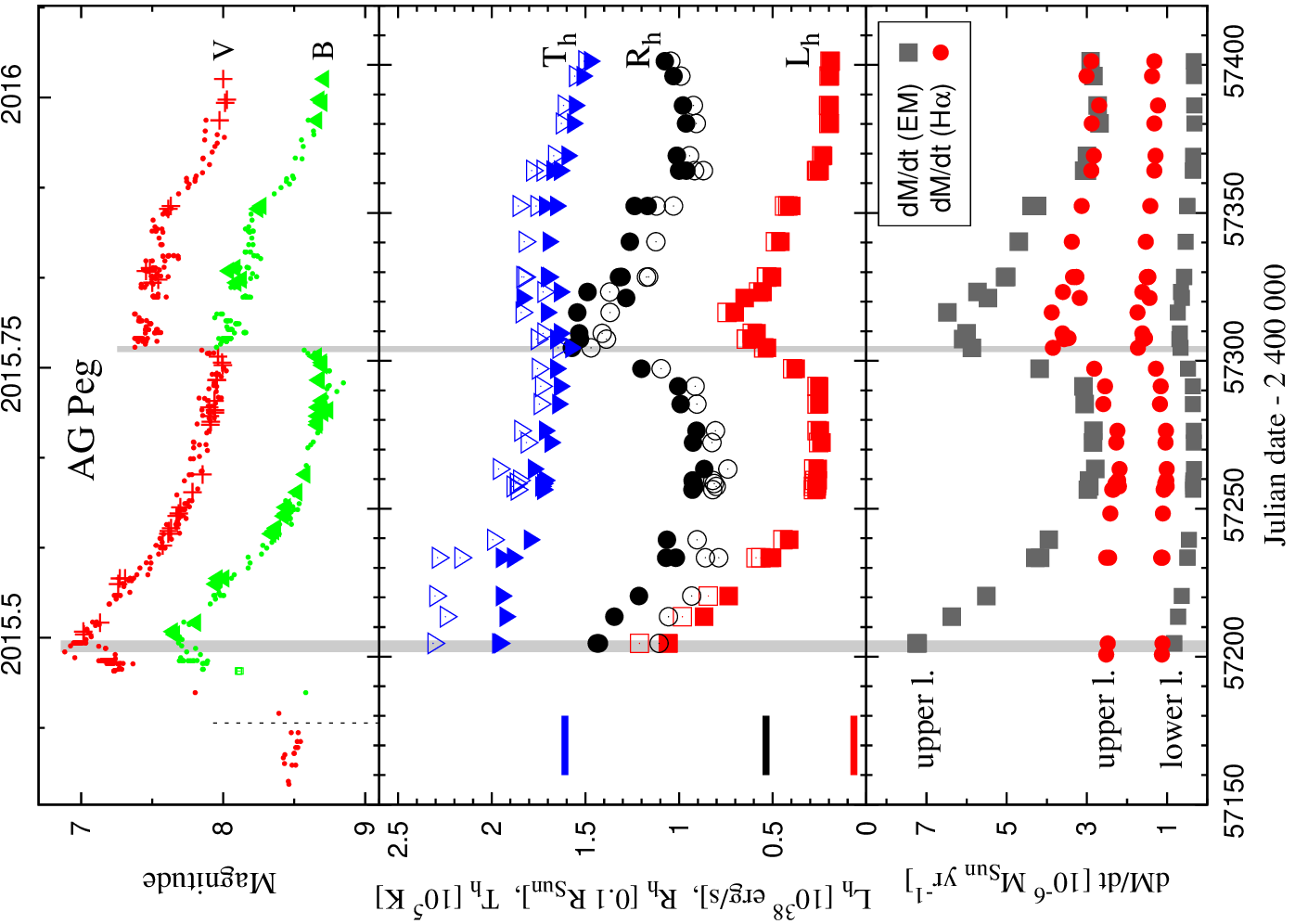}}
\end{center}
\caption[]{
Evolution of the parameters $L_{\rm h}$, $R_{\rm h}$, 
$T_{\rm h}$ (middle), and the mass-loss rate 
$\dot M_{\rm h}$ from the hot component (bottom; the upper 
and lower limits are denoted by `upper l.' and `lower l.'). 
The dotted and grey lines mark the beginning of the explosion 
(June 5.10$\pm 3.3$) and the maxima in the \textsl{LC} 
(June 30.0$\pm 2.0$, Oct. 8.5$\pm 0.5$). Filled and open 
symbols correspond to $T_{\rm h}$ from Eqs.~(\ref{eq:heh2}) 
and (\ref{eq:tcross}). 
Data are listed in Table~\ref{tab:lrt}. 
          }
\label{fig:lrt}
\end{figure}

The presence of both high and low ionisation species in the 
spectrum of AG~Peg (e.g. \ion{O}{vi}, [\ion{Fe}{vii}], 
[\ion{O}{iii}], \ion{N}{iii}, \ion{He}{ii}, \ion{H}{i}, 
\ion{He}{i}, \ion{Fe}{ii}, [\ion{O}{i}]\,$\lambda 6300$)
suggests that its nebula is rather ionisation-bounded, 
that is, the stellar radiation above $\nu_0$ (13.6\,eV) is absorbed 
within the nebula \citep[see][ and references therein]{kj89}. 
The right side of Eq.~(\ref{eq:heh}) assumes that the lines 
are optically thin. This, however, need not to be satisfied 
in the very dense symbiotic nebulae (e.g. the decrement of 
hydrogen Balmer lines usually departs from theoretical 
prediction). 
However, the ratio of the fluxes lowers possible deviation from 
the optically thin case. 
Applying Eq.~(\ref{eq:heh}) to \ion{He}{ii}($\lambda 4686$)/\hb\ 
flux ratios in the spectrum of AG~Peg (Table~6), we used 
recombination coefficients from \cite{hs87} for $T_{\rm e}$ = 
20000\,K (Table~\ref{tab:sed}) and electron concentration, 
$n_{\rm e} = 10^{10}$\cmt\ \citep[see][]{sk+11}: 
$\alpha_{\rm B}({\rm H}^0) = 1.53\times 10^{-13}$, 
$\alpha({\rm H}\beta) = 1.74\times 10^{-14}$, 
$\alpha_{\rm B}({\rm He}^{+}) = 1.02\times 10^{-12}$ and 
$\alpha_{4686} = 1.48\times 10^{-13}$\,cm$^3$\,s$^{-1}$. 
According to possible large uncertainties of $T_{\rm e}$ and 
$n_{\rm e}$, it is important to note that the ratio of 
$\alpha_{\rm B}$ to $\alpha$(line) depends on $T_{\rm e}$ and 
$n_{\rm e}$ only marginally. Equation~(\ref{eq:heh}) 
then reads as, 
\begin{equation}
 \frac{Q(4\nu_0,\infty)}{Q(\nu_0,4\nu_0)} = 
  0.754\times \frac{F_{4686}}{F_{{\rm H}\beta}}, 
\label{eq:heh2}
\end{equation}
where the number of quanta on the left side was calculated 
for Planck's function. Resulting $T_{\rm h}$ runs from 
$\la 2\times 10^5$\,K at the beginning of the outburst 
to $\sim 1.5\times 10^5$\,K in January 2016 
(see Fig.~\ref{fig:lrt} and Table~\ref{tab:lrt}). 

During the quiescent phase, the nebula is only partly 
ionisation-bounded, that is, a fraction of ionising photons 
escapes the nebula. 
As the H$^{+}$ zone is always more open than the He$^{+2}$ zone 
\citep[in the sense of the STB model, see Fig.~1 of][]{nv87}, 
the relative fraction of the non-absorbed photons capable of 
ionising hydrogen will always be larger than that capable of 
ionising He$^{+}$ ions. As a result, the observed ratio 
$(F_{4686}/F_{\rm H\beta})_{\rm obs.} > F_{4686}/F_{\rm H\beta}$ 
in Eq.~(\ref{eq:heh}). 
Therefore, during quiescent phases, the \ion{He}{ii}/\hb\ method 
indicates only the upper limit of $T_{\rm h}$. 
In addition, $F_{\rm H\beta}$ and thus $T_{\rm h}$ changes with 
orbital phase \citep[see][]{kenyon+01}, because the nebula is 
partially optically thick (Sect.~\ref{sec:phot}). 
As Eq.~(\ref{eq:heh}) assumes that both the lines are optically 
thin, maximum fluxes at $\varphi\sim 0.5$ (WD in front) correspond 
to the most appropriate upper limit of $T_{\rm h}$. Our values 
of 156--168\,kK (Table~\ref{tab:lrt}) that we obtained during the 
quiescent phase around $\varphi = 0.5$, thus represent 
an upper limit of $T_{\rm h}$. Therefore, in modelling the SED 
during the quiescent phase (Sect.~\ref{sec:sed}) we adopted 
a lower value of $T_{\rm h} = 150000$\,K. 

\cite{kj89} used the Ambartsumyan's approach and calculated 
the Zanstra \ion{H}{i} and \ion{He}{ii} temperatures. In an 
iteration process, forcing agreement between all temperatures, 
they derived the so-called `crossover temperature', 
$T_{\rm cross}$. For $0.08 < F_{4686}/F_{{\rm H}\beta} < 1$ 
they determined its polynomial approximation as 
%
\begin{eqnarray}
 \log(T_{\rm cross}) = 4.905+1.11162\times 10^{-2} I_{\rm c}
\nonumber \\
                      -1.10692\times 10^{-4} I_{\rm c}^2
\nonumber \\
                      +6.20572\times 10^{-7} I_{\rm c}^3,
\label{eq:tcross}
\end{eqnarray}
%
where $I_{\rm c} = 100\times F_{4686}/F_{{\rm H}\beta}$. 
In this way determined values of $T_{\rm h}$ were higher than 
those obtained by the \ion{He}{ii}($\lambda 4686$)/\hb\ method. 
A larger difference of 15--18\% was indicated only at the 
beginning of the outburst (see Fig.~\ref{fig:lrt}). 

Independently, a high $T_{\rm h}$ was suggested by \cite{luna+15}, 
who detected a supersoft black-body type component in the X-ray 
spectrum of AG~Peg with kT$\sim 0.02$\,keV by the Swift satellite 
at the end of June 2015. However, \cite{ramsay+16} did not mention 
this result, because they couldn't find strong evidence to confirm 
the presence of this supersoft component (Luna 2016, private 
communication). 
%
%

\subsubsection{Luminosity and radius of the ionising source}
\label{sec:lhrh}

Having independently determined values of $T_{\rm h}$ and 
$EM$, we can estimate the luminosity $L_{\rm h}$ of the hot 
component, and thus its effective radius, $R_{\rm h}$ under 
the assumption that the total flux of its ionising photons, 
$Q(\nu_0,\infty)$, is balanced by the total rate of 
recombinations within the ionised volume $V$, that is, 
\begin{equation}
 Q(\nu_0,\infty) = 
              \alpha_{\rm B}({\rm H}^0,T_{\rm e})
              \!\int_{V} \!\!n_{+}(r)n_{\rm e}(r)\,{\rm d}V 
            = \alpha_{\rm B}({\rm H}^0,T_{\rm e}) EM. 
\label{eq:lph}
\end{equation}
In this way the ionising photons are converted into the diffuse 
radiation, which we indicate in the spectrum as the nebular 
continuum and emission lines. 
Its amount in the continuum is given 
by the value of \textsl{EM,} which is determined by the 
$F_{\rm n}(\lambda)$ component in the model SED as 
(see Sect.~\ref{sec:sed}), 
\begin{equation}
 4\pi d^2 F_{\rm n}(\lambda) =
   \varepsilon_{\lambda}({\rm H}^0,T_{\rm e})
               \!\int_{V}\!\! n_{+}n_{\rm e}\,{\rm d}V = 
   \varepsilon_{\lambda}({\rm H}^0,T_{\rm e}) EM .
\label{eq:fneb}
\end{equation}
This includes the kinetic energy of free electrons released 
into the continuum via f--f and f--b transitions. 
Using the expression for $Q(\nu_0,\infty)$ given by Eq.~(11) 
of \cite{sk01}, the luminosity $L_{\rm h}$ of the ionising 
source, which generates the observed \textsl{EM}, can be 
expressed as 
\begin{equation}
 L_{\rm h} = \alpha_{\rm B}({\rm H}^0,T_{\rm e})\,\textsl{EM}
                \frac{\sigma T_{\rm h}^{4}}{f(T_{\rm h})},
\label{eq:lth}
\end{equation}
where the function
\begin{equation}
f(T_{\rm h}) = \frac{\pi}{hc}\int^{\rm 912\AA}_{0}\!\!\!
               \lambda\, B_{\lambda}(T_{\rm h})\,\rm d\lambda\label{eq:fth}
\end{equation}
determines the flux of ionising photons emitted by 1\,cm$^2$ 
area of the hot component photosphere (cm$^{-2}$\,s$^{-1}$). 
Finally, the effective radius of the hot component, $R_{\rm h}$, 
is determined according to Stefan-Boltzmann law as 
\begin{equation}
  R_{\rm h} = \sqrt{\frac{L_{\rm h}}{4\pi \sigma T_{\rm h}^{4}}},
\label{eqn:R}
\end{equation}
where $\sigma$ is the Stefan-Boltzmann constant. 

To determine $L_{\rm h}$ using $T_{\rm h}$ and \textsl{EM}, 
we interpolated the original values of \textsl{EM} obtained 
from the low-resolution spectra to dates of medium-resolution 
spectra (the source of $T_{\rm h}$, see Fig.~\ref{fig:param}). 
Temporal evolution of parameters $T_{\rm h}$, $R_{\rm h}$, and 
$L_{\rm h}$ along the outburst is shown in Fig.~\ref{fig:lrt}. 
Corresponding data are in Table~\ref{tab:lrt}. 

Maximum values of all parameters correspond to the maximum of 
the star's brightness, because the nebular emission dominated 
the optical. As this component was strong and variable 
(Fig.~\ref{fig:sedopt}), the evolution in the \textsl{EM} thus 
governed the profile of both the \textsl{LC} and other dependent 
parameters in the course of the outburst 
(see Sect.~\ref{sec:dis} in detail). 

Finally, determination of the fundamental parameters, 
$L_{\rm h}$, $R_{\rm h}$ and $T_{\rm h}$ (Sects.~\ref{sec:th} 
and \ref{sec:lhrh}) was made under the assumption that the ionising 
source radiates as a black body. Reliability of this approach 
is discussed in Appendix~B. 

\subsubsection{Mass-loss rate from the hot component}
\label{sec:mdot}

It was shown that the nebular emission during active phases of 
symbiotic stars is generated by the ionised stellar wind from 
their hot components \citep[][]{sk06}. Hence, the significant 
increase of the nebular emission in the spectrum of AG~Peg, in 
both the continuum and lines (see Figs.~\ref{fig:param} and 
\ref{fig:broad}), suggests a significant increase of the 
mass-loss rate, $\dot M_{\rm h}$, from its hot component in 
the form of ionised wind. 
Below we estimate $\dot M_{\rm h}$ from the nebular continuum 
and the \ha\ emission. 

We assume a spherically symmetric wind around the hot component 
with the beginning at the radial distance from the WD centre 
$r = R_0$, that becomes optically thin at $r = R_{\rm in}$. 
Further, the particle velocity of the wind satisfies 
the $\beta$-law wind as introduced by \cite{lamcass99}, 
\begin{equation}
  v(r) = v_{\infty}\left(1-\frac{b R_0}{r}\right)^{\beta},
\label{eq:betalaw}
\end{equation}
where $\beta$ determines the acceleration of the wind, $v_{\infty}$ 
is its terminal velocity, and the parameter $b$ is given by 
\begin{equation}
  b = 1-\left(\frac{a}{v_{\infty}}\right)^{1/\beta},
\label{eq:a}
\end{equation}
where $a$ is the initial velocity of the wind at its origin. 

{\sf Mass-loss rate from the \textsl{EM}}. 
For the ionised wind, as specified above, we can express 
its \textsl{EM} as (see Appendix~C), 
\begin{equation}
\textsl{EM} = \frac{4\pi}{(4\pi\mu m_{\rm H})^2}
              \Big(\frac{\dot M_{\rm h}}{v_{\infty}}\Big)^{2}
              \frac{1}{b R_0 (1-2\beta)}
              \Big[1-\Big(1-\frac{b R_0}{R_{\rm in}}
              \Big)^{1-2\beta}\Big],
\label{eq:em}
\end{equation}
where $\mu$ is the mean molecular weight and $m_{\rm H}$ is 
the mass of the hydrogen atom. Extension of \ha\ wings suggests 
$v_{\infty} \sim 2000$\kms\ (Fig.~\ref{fig:param}). According 
to a model of the broad \ha\ wings in the spectra of symbiotic 
binaries \citep[see][]{sk06}, we adopted $\beta \sim 1.7$ and, 
according to \cite{cs12}, we adopted $a$ = 50\kms. 
We assume that the wind becomes optically thin in the continuum 
at the WD's pseudophotosphere, that is, $R_{\rm in} = R_{\rm h}$. 
Following the theory of the so-called optically thick wind 
in nova outbursts, in which the matter is accelerated deep 
inside the photosphere \citep[e.g.][]{k+h94}, we can estimate 
the lower and the upper limit of the mass-loss rate as follows. 
\begin{itemize}
\item 
If the wind begins at the WD's pseudophotosphere 
(= the optically thick/thin interface of the wind), that is, 
$R_{\rm 0} = R_{\rm h}$, we obtain the lower limit of 
$\dot M_{\rm h}$ (see Eq.~(\ref{eq:em})). The measured values 
of \textsl{EM} and $R_{\rm h}$ imply 
$\dot M_{\rm h} = 3.3-8.3\times 10^{-7}$\myr. 
\item
If the wind begins at the WD's surface, that is, 
$R_{\rm 0} = R_{\rm WD}$, Eq.~(\ref{eq:em}) provides the upper 
limit of $\dot M_{\rm h}$. Our measurements and 
$R_{\rm WD} \equiv 0.01$\ro\ correspond to 
$\dot M_{\rm h} = 2.7-7.2\times 10^{-6}$\myr. 
\end{itemize}
In the real case, the wind probably begins somewhere between 
$R_{\rm WD}$ and $R_{\rm h}$. For example, setting 
$R_{\rm 0}$ = 0.06\ro\ results in 
$\dot M_{\rm h} = 1.4-5.1\times 10^{-6}$\myr. 
Table~\ref{tab:lrt} presents the average of the upper and lower 
limit of $\dot M_{\rm h}$. We note that the method is not 
applicable for the quiescent phase, because the corresponding 
\textsl{EM} represents the nebular emission from the 
ionised wind from the giant. 

{\sf Mass-loss rate from the \ha\ flux}.
Using the same assumptions as for the continuum radiation, 
the measured flux in the \ha\ line, $F_{{\rm H}\alpha}$, allows 
us to express its luminosity as 
\begin{equation}
 4\pi d^2 F_{{\rm H}\alpha} = h\nu({\rm H}\alpha)
          \alpha({\rm H}\alpha,T_{\rm e})
          \int_{R_{\rm in}}^{\infty}\!\! n_{\rm e}(r)
                               n_{\rm p}(r)\,{\rm d}V, 
\label{eq:lha}
\end{equation}
where $\alpha({\rm H}\alpha,T_{\rm e})$ is the effective 
recombination coefficient for the \ha\ transition, and $n_{\rm e}$ 
and $n_{\rm p}$ is the concentration of electrons and protons in 
the spherical H$^{+}$ zone with the inner radius $r = R_{\rm in}$, 
at which the wind becomes optically thin in the \ha\ line. 
This fundamental condition was treated by \cite{leitherer88} 
for stellar winds of OB stars. He found that winds of O stars 
are optically thin from a distance $r \sim 1.5 R_{\star}$. 
Also, modelling the broad \ha\ wings in the spectra of symbiotic 
stars suggested the validity of the optically thin regime from 
about 1.2 to 1.5\,$R_{\rm h}$ \citep[][]{sk06}. Therefore, we 
adopted $R_{\rm in} = 1.5\times R_{\rm h}$. 

Applying Eq.~(\ref{eq:lha}), with the expression for the 
\textsl{EM} given by Eq.~(\ref{eq:em}), to our values of 
$F_{{\rm H}\alpha}$ and $R_{\rm h}$ 
(Tables~\ref{tab:lrt} and 6), 
we obtain $\dot M_{\rm h} = 1.0-1.6\times 10^{-6}$\myr\ for 
$R_{\rm 0} = R_{\rm h}$, and $2.2-3.6\times 10^{-6}$\myr\ for 
$R_{\rm 0} = 0.01$\ro. We used the volume emission coefficient 
$h\nu({\rm H}\alpha)\alpha({\rm H}\alpha, 20000\,{\rm K}) 
= 1.83\times 10^{-25}$\,$\rm erg\,cm^3 s^{-1}$. 
%
%
%
\begin{table}
\caption[]{Parameters of the hot component, 
           $L_{\rm h}$ (10$^{37}$\es), 
           $R_{\rm h}$ (\ro), 
           $T_{\rm h}$ (10$^{5}$\,K) and 
           $\dot{M}_{\rm h}$ (10$^{-6}$\myr) 
(see Sects.~\ref{sec:th}, \ref{sec:lhrh} and \ref{sec:mdot}).}
\begin{center}
\begin{tabular}{ccccc}
\hline
\hline
Date            & 
$L_{\rm h}$     & 
$R_{\rm h}$     &
$T_{\rm h}^{~a)}$ &
$\dot{M}_{\rm h}^{~b)}$ \\
\hline
         \multicolumn{5}{c}{Quiescent phase} \\
\hline
 2013/08/11.900 & 0.6$\pm 0.1$&  0.056$\pm 0.003$&  1.56$\pm 0.06$&  -- \\
 2013/09/04.855 & 0.6$\pm 0.1$&  0.058$\pm 0.003$&  1.54$\pm 0.06$&  -- \\
 2013/09/27.978 & 0.7$\pm 0.1$&  0.054$\pm 0.003$&  1.60$\pm 0.06$&  -- \\
 2013/10/24.874 & 0.7$\pm 0.1$&  0.053$\pm 0.003$&  1.62$\pm 0.06$&  -- \\
 2013/10/26.793 & 0.7$\pm 0.1$&  0.051$\pm 0.003$&  1.66$\pm 0.07$&  -- \\
 2013/12/17.716 & 0.7$\pm 0.1$&  0.050$\pm 0.003$&  1.68$\pm 0.07$&  -- \\
\hline
          \multicolumn{5}{c}{Active phase} \\ 
\hline
 2015/07/01.011 & 10.6$\pm 1.0$&  0.143$\pm 0.009$&  1.97$\pm 0.08$&  4.0 \\
 2015/07/09.999 &  8.7$\pm 0.8$&  0.135$\pm 0.008$&  1.93$\pm 0.08$&  3.5 \\
 2015/07/17.004 &  7.4$\pm 0.7$&  0.121$\pm 0.007$&  1.95$\pm 0.08$&  3.1 \\
 2015/07/29.942 &  5.0$\pm 0.5$&  0.107$\pm 0.006$&  1.89$\pm 0.08$&  2.4 \\
 2015/07/29.942 &  5.2$\pm 0.5$&  0.102$\pm 0.006$&  1.95$\pm 0.08$&  2.3 \\
 2015/08/04.971 &  4.1$\pm 0.4$&  0.106$\pm 0.006$&  1.80$\pm 0.07$&  2.2 \\
 2015/08/21.896 &  2.7$\pm 0.2$&  0.093$\pm 0.006$&  1.73$\pm 0.07$&  1.7 \\
 2015/08/23.002 &  2.7$\pm 0.2$&  0.092$\pm 0.005$&  1.74$\pm 0.07$&  1.6 \\
 2015/08/23.983 &  2.6$\pm 0.2$&  0.092$\pm 0.006$&  1.73$\pm 0.07$&  1.6 \\
 2015/08/25.039 &  2.6$\pm 0.2$&  0.093$\pm 0.006$&  1.72$\pm 0.07$&  1.6 \\
 2015/08/28.927 &  2.6$\pm 0.2$&  0.087$\pm 0.005$&  1.78$\pm 0.07$&  1.6 \\
 2015/09/06.883 &  2.4$\pm 0.2$&  0.092$\pm 0.006$&  1.69$\pm 0.07$&  1.6 \\
 2015/09/10.852 &  2.5$\pm 0.2$&  0.091$\pm 0.005$&  1.72$\pm 0.07$&  1.6 \\
 2015/09/19.875 &  2.5$\pm 0.2$&  0.099$\pm 0.006$&  1.65$\pm 0.07$&  1.7 \\
 2015/09/25.837 &  2.5$\pm 0.2$&  0.101$\pm 0.006$&  1.64$\pm 0.07$&  1.7 \\
 2015/10/01.825 &  3.8$\pm 0.3$&  0.120$\pm 0.007$&  1.66$\pm 0.07$&  2.3 \\
 2015/10/08.849 &  5.3$\pm 0.5$&  0.157$\pm 0.009$&  1.58$\pm 0.06$&  3.3 \\
 2015/10/11.779 &  6.1$\pm 0.6$&  0.153$\pm 0.009$&  1.66$\pm 0.07$&  3.4 \\
 2015/10/13.777 &  5.9$\pm 0.5$&  0.153$\pm 0.009$&  1.64$\pm 0.07$&  3.3 \\
 2015/10/20.830 &  7.0$\pm 0.6$&  0.154$\pm 0.009$&  1.71$\pm 0.07$&  3.6 \\
 2015/10/25.747 &  6.5$\pm 0.6$&  0.128$\pm 0.008$&  1.84$\pm 0.07$&  3.0 \\
 2015/10/27.780 &  5.5$\pm 0.5$&  0.149$\pm 0.009$&  1.64$\pm 0.07$&  3.2 \\
 2015/11/01.771 &  5.0$\pm 0.5$&  0.132$\pm 0.008$&  1.70$\pm 0.07$&  2.8 \\
 2015/11/01.807 &  5.0$\pm 0.5$&  0.130$\pm 0.008$&  1.72$\pm 0.07$&  2.7 \\
 2015/11/01.840 &  5.0$\pm 0.5$&  0.131$\pm 0.008$&  1.71$\pm 0.07$&  2.8 \\
 2015/11/13.749 &  4.6$\pm 0.4$&  0.126$\pm 0.008$&  1.70$\pm 0.07$&  2.6 \\
 2015/11/25.801 &  4.1$\pm 0.4$&  0.117$\pm 0.007$&  1.72$\pm 0.07$&  2.4 \\
 2015/12/07.805 &  2.5$\pm 0.2$&  0.100$\pm 0.006$&  1.64$\pm 0.07$&  1.7 \\
 2015/12/12.803 &  2.3$\pm 0.2$&  0.101$\pm 0.006$&  1.60$\pm 0.06$&  1.7 \\
 2015/12/23.735 &  1.9$\pm 0.2$&  0.096$\pm 0.006$&  1.57$\pm 0.06$&  1.5 \\
 2015/12/29.761 &  2.0$\pm 0.2$&  0.098$\pm 0.006$&  1.56$\pm 0.06$&  1.5 \\
 2016/01/08.720 &  2.0$\pm 0.2$&  0.103$\pm 0.006$&  1.52$\pm 0.06$&  1.6 \\
 2016/01/13.723 &  1.9$\pm 0.2$&  0.108$\pm 0.006$&  1.48$\pm 0.06$&  1.6 \\
\hline
\end{tabular}
\end{center}
{\bf Notes.} 
$^{a)}$~ According to Eq.~(\ref{eq:heh2}), 
$^{b)}$~ the method is not applicable for quiescent phase. 
\label{tab:lrt}
\end{table}  
%
%
%
\begin{figure}
\begin{center}
\resizebox{\hsize}{!}{\includegraphics[angle=-90]{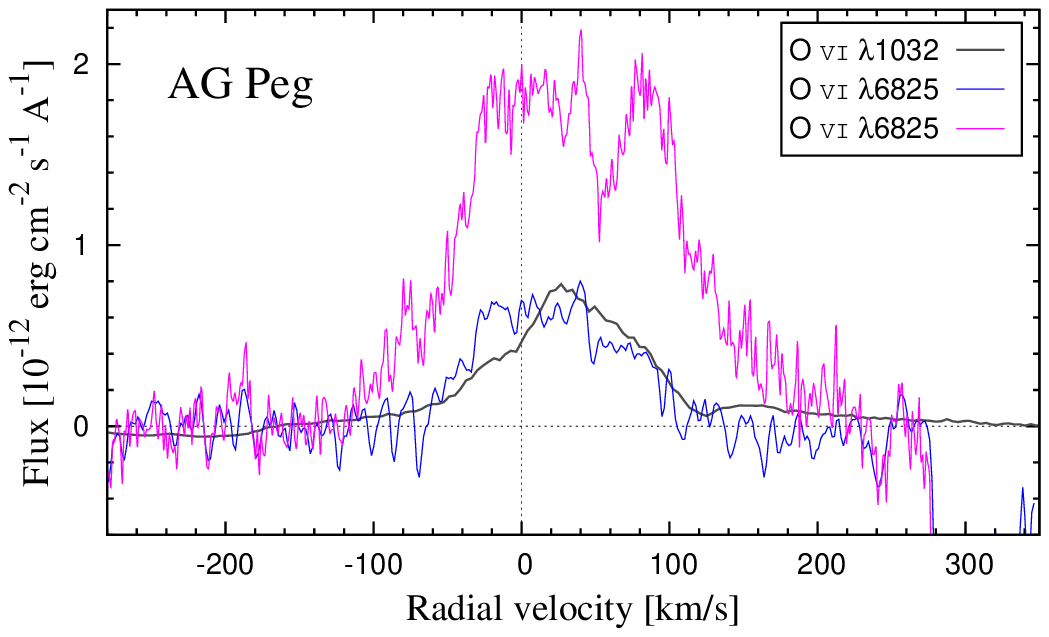}}
\end{center}
\caption[]{
Example of the Raman scattered \ion{O}{vi} 6825\,\AA\ line 
during the quiescent phase (blue line; August 29.945, 2013) 
and the outburst (magenta line; August 28.927, 2015). 
Flux of the original \ion{O}{vi} 1032\,\AA\ line (grey line; 
June 5.618, 2003) was scaled to that of the Raman line 
from quiescence. 
The radial velocity scale corresponds to the velocity space of 
the original \ion{O}{vi} line for the systemic velocity of 
-16\kms\ \citep[][]{fekel+00} and the wavelength of the Raman 
transition, 6825.44\,\AA\ \citep[][]{schmid89}. 
          }
\label{fig:ram1}
\end{figure}
%
%
%
\begin{figure}
\begin{center}
\resizebox{\hsize}{!}{\includegraphics[angle=-90]{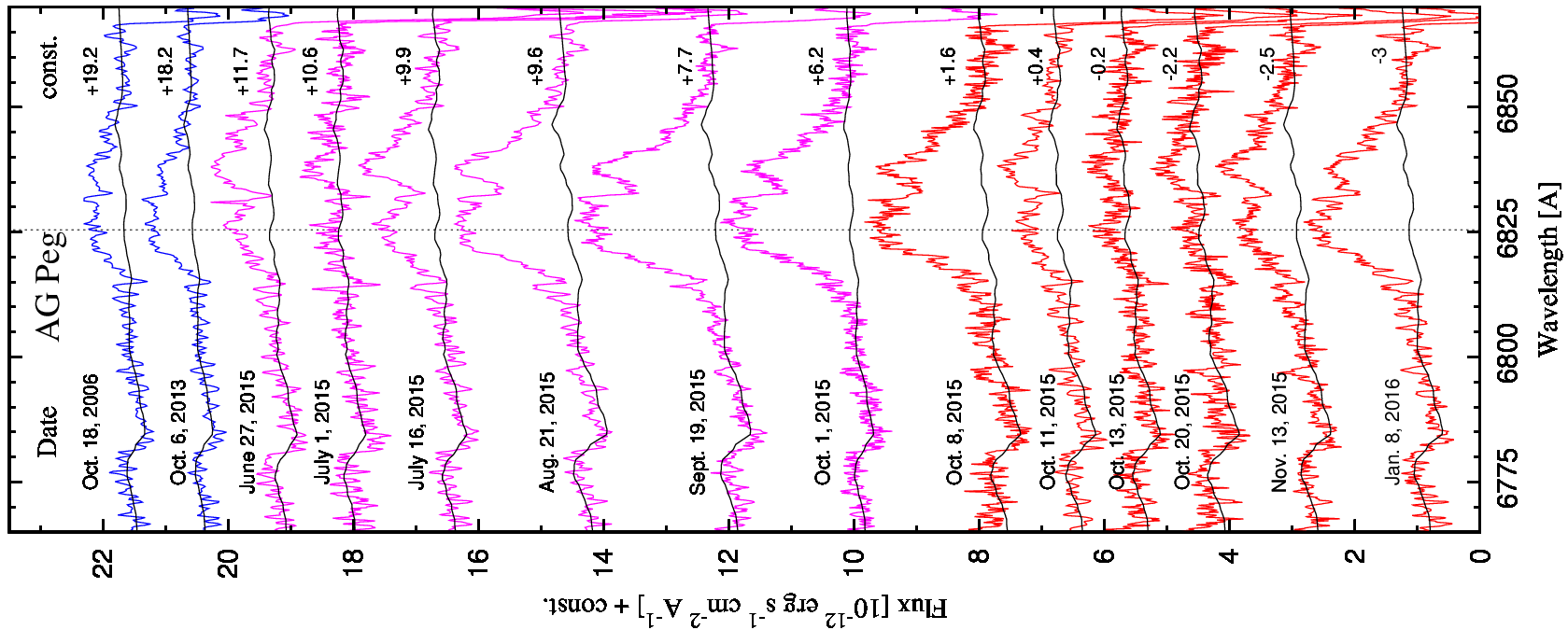}}
\end{center}   
\caption[]{
Evolution of the Raman-scattered \ion{O}{vi} 6825\,\AA\ line prior 
to (blue lines) and along the 2015 outburst: at/after the first 
maximum on June 30th (magenta lines) and the second one on October 
8th (red lines). The black lines represent the continuum. 
Numbers on the right side mark the shift of the local 
continuum for a better visualisation. 
          }
\label{fig:ram2}
\end{figure}

\subsubsection{Raman-scattered \ion{O}{vi} lines}
\label{sec:ovi}
\cite{schmid89} first identified broad emission bands at 
6830\,\AA\ and 7088\,\AA\ in symbiotic stars as a result of 
Raman-scattering of the \ion{O}{vi} 1032\,\AA\ and 1038\,\AA\ 
line photons on neutral hydrogen atoms. The Raman scattered 
\ion{O}{vi} 
features were also observed in the spectrum of 
AG~Peg \citep[][ and this paper]{schmid+99}. 

The last observation of the \ion{O}{vi}\,1032 and 1038\,\AA\ 
doublet was obtained with the \textsl{FUSE} satellite during 
quiescent phase, on June 5.618, 2003, near to the inferior 
conjunction of the giant ($\varphi\sim 0.9$, see Fig.~\ref{fig:ulc}). 
We estimated the observed flux of the \ion{O}{vi} 1032\,\AA\ 
line to $\sim 96\times 10^{-12}$\ecs, which corresponds to 
the dereddened flux of $\sim 384\times 10^{-12}$\ecs\ (see 
Appendix~B). 
During the quiescent phase (our spectra from 2006 and 2013), 
the $FWZI$ of the Raman \ion{O}{vi} 6830\,\AA\ line was 
25--30\,\AA, and its dereddened flux of 
12.5$\times 10^{-12}$\ecs. These quantities give the efficiency 
of the Raman scattering $\eta$ (= $6.614 \times 
F_{6825}/F_{1032}$) of $\sim 21.5$\%. 
The width of the Raman line was comparable with that measured 
during 1993 and 1996, whereas its dereddened flux increased by 
a factor of $\sim 5-7$ and the scattering efficiency by a factor 
of $\sim 3-5$ \citep[see Table~7 of][]{schmid+99}. 
The increase of the Raman scattering efficiency was probably 
caused by an enlargement of the neutral hydrogen region on 
the sky of the \ion{O}{vi} zone located around the WD. This is 
according to the basic ionisation structure of symbiotic stars 
during quiescence, where the angular size of the neutral 
\ion{H}{i} region, as seen from the WD, depends on the flux 
of ionising photons from the WD and on the flux of neutral hydrogen 
from the giant (STB). Since the hot component 
luminosity was gradually declining (see Sect.~\ref{sec:intr}) 
and the mass-loss rate from the giant was presumably constant, 
the opening angle of the \ion{H}{i} region, and thus the Raman 
conversion efficiency, were increasing. 
Also, a higher efficiency derived from the 2003 \textsl{FUSE} 
spectrum could be caused by measuring this spectrum at 
the orbital phase $\varphi \sim 0.9$, where a fraction of the 
original \ion{O}{vi} photons can be attenuated by the neutral 
wind from the giant. 

Figure~\ref{fig:ram1} compares profiles of the original 
\ion{O}{vi} 1032\,\AA\ line and its Raman scattered fraction. 
During quiescence, both profiles are nearly overlapped, except 
a blueward-shifted component in the Raman profile between -30 
and 0\,\kms. According to \cite{schmid+99} this emission can 
be produced by the densest \ion{H}{i} atoms of the giant's wind 
around the binary axis, that move against the \ion{O}{vi} region. 
For a steep wind velocity profile of giants in symbiotic 
binaries, as suggested by \cite{knill+93} and recently applied 
by \cite{shag+16} (see their Fig.~5), the edge of this 
adjacent component at $\sim -30$\,\kms\ can be close to 
the terminal velocity of the giant's wind. 

During the outburst, significant variations in the Raman 
6825\,\AA\ line profile were measured. Figure~\ref{fig:param} 
(bottom panel) shows evolution of its fluxes, whereas 
Figure~\ref{fig:ram2} compares its profiles along the outburst. 
Throughout the whole outburst, a redward-shifted component 
around 6838\,\AA\ developed -- analogous to Z~And during its 
2000--2003 and 2006--2007 
active phases \citep[see Fig~4 of][]{sk+06} and Fig.~5 of \cite{sk+09b}. 
Except for minimum fluxes (around July 1 and October 11) 
the $FWZI$ of the Raman line enlarged to $\sim$\,40\,\AA, and 
its flux increased by a factor of 4--5 relative to that from 
the preceding quiescence. 
A minimum flux of the Raman line was observed just after 
the optical maxima (Fig.~\ref{fig:param}). 
Finally, a few faint absorption lines of the giant spectrum 
can be recognised within the Raman 6825\,\AA\ feature (e.g. at 
$\lambda$6807.2, $\lambda$6810.3, $\lambda$6812.5, 
$\lambda$6815.1, $\lambda$6824.8, $\lambda$6832.7, 
$\lambda$6840.0, $\lambda$6846.9 and $\lambda$6850.3). 
However, their influence on major features of the Raman 
profile is negligible. 
%

\begin{figure}
\begin{center}
\resizebox{\hsize}{!}{\includegraphics[angle=-90]{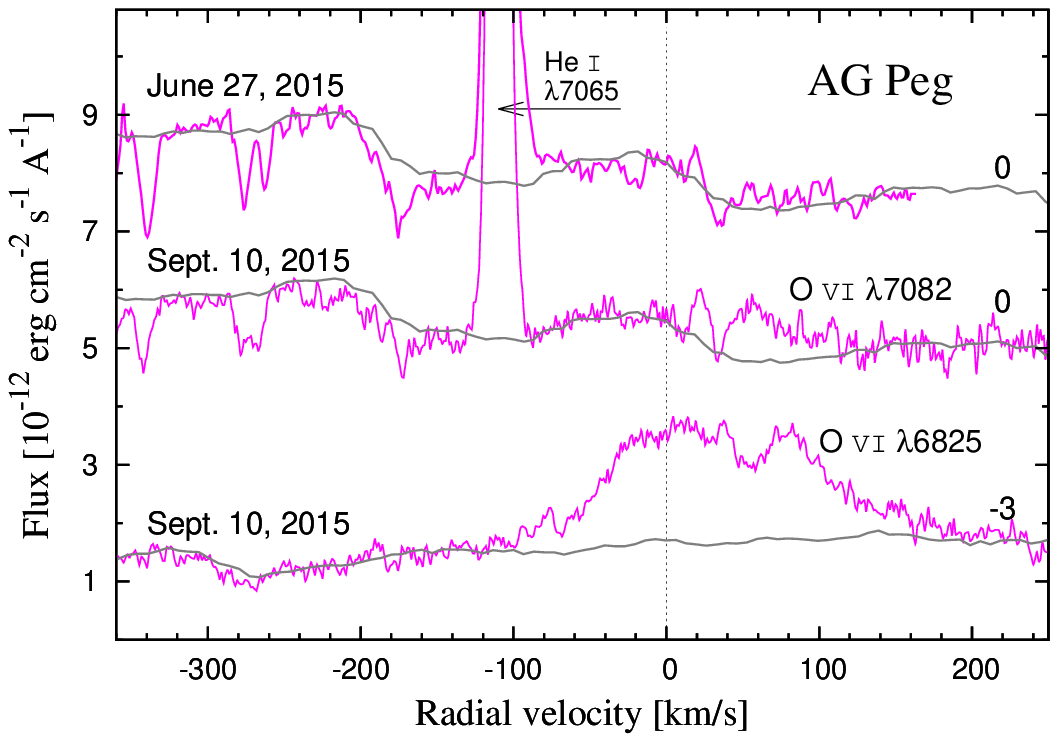}}
\end{center}
\caption[]{
Example of the Raman scattered \ion{O}{vi} 7082\,\AA\ line 
around its minimum (top) and maximum flux (middle). 
Compared is the Raman \ion{O}{vi} 6825\,\AA\ line (bottom). 
The grey line represents the continuum according to model 
SEDs. Radial velocity scale corresponds to the systemic 
velocity space of the original \ion{O}{vi}\,1031.928 or 
1037.618\,\AA\ line and wavelengths of their Raman 
transitions at 6825.44 or 7082.40\,\AA. 
          }
\label{fig:ram3}
\end{figure}

Concerning the Raman \ion{O}{vi} 7082\,\AA\ line, its flux 
was a factor of $\sim 10$ smaller than that of the Raman 6825\,\AA\ 
line during their best visibility (mid August -- end of September, 
2015). Also, profiles of both Raman lines were not comparable. 
For example, the whole blue part of the 7082\,\AA\ profile was difficult 
to detect (see Fig.~\ref{fig:ram3}). This was probably because of 
its weakness, and partly because the stellar spectrum is very 
structured around this wavelength. For example, the central part 
of the profile is strongly affected by an absorption at 
$\sim 7087.5$\,\AA. 
During the minimum fluxes of $\lambda 6825$ Raman line, the 
$\lambda 7082$ Raman emission was not recognisable in our spectra. 

Observed variations and possible origin of the Raman 6825\,\AA\
emission during the outburst of AG~Peg is discussed in 
Sect.~\ref{dis:ram}. 
%
%
%
%

\section{Discussion}
\label{sec:dis}

\subsection{The hot component temperature}
\label{dis:th}
Following ionisation structure of hot components in symbiotic 
binaries during active phases, the high temperature of the WD's 
pseudophotosphere results from a low orbital inclination 
of AG~Peg. 
In this case the observer can directly see the hot component 
in contrast to systems with high orbital inclination, when 
its hard radiation is blocked by the optically thick disk-like 
pseudophotosphere, which develops during outbursts 
\citep[][]{sk05b}. Our values of 150 -- 230\,kK are comparable 
to those indicated in other symbiotic systems with the observed 
supersoft X-ray and far-UV fluxes generated by the WD's 
pseudophotosphere. For example, AG~Dra 
\citep[115--180\,kK,][]{greiner+97,sk+09a}, 
RR~Tel \citep[150--180\,kK,][]{g-r+13}, and 
LIN~358 \citep[250$\pm 10$\,kK,][]{sk15}. 

To estimate $T_{\rm h}$, \cite{ramsay+16} used only equivalent widths 
of \ion{He}{ii}\,4686\,\AA\ and \hb\ lines, instead of their fluxes. 
Despite variation of the continuum level by an amplitude of 
$\sim 1$\,mag in the $BV$ passbands over the course of the 
outburst, the authors obtained values of $T_{\rm h}$ between 
$\sim$1.8 and $\sim 1.4\times 10^5$\,K, which are comparable 
with our quantities. Similar temperatures were also achieved 
by \cite{tomov+16} using the same approach to their 
high-resolution spectra. 
This is the result of a close vicinity of both the lines in the 
spectrum ($\Delta\lambda \sim 175$\,\AA) and the method 
that uses their ratio. 

\subsection{The hot component luminosity}
\label{dis:lh}
The high luminosity of the hot component in AG~Peg of a few times 
$10^{37}$\es, with a maximum of $\sim 10^{38}$\es, is a natural 
consequence of the high value of the \textsl{EM}, measured during 
the 2015 outburst. For example, 
\textsl{EM} = 5--9$\times 10^{60}$\cmt\ requires the rate of 
hydrogen-ionising photons, 
$Q(\nu_0,\infty) = \alpha_{\rm B}(T_{\rm e})\times \textsl{EM} 
\sim 10^{48}$\,s$^{-1}$ for 
$\alpha_{\rm B}(20000) = 1.53\times 10^{-13}$\,cm$^{3}$\,s$^{-1}$ 
(see Sect.~\ref{sec:th}). 
Such a high rate of $Q(\nu_0,\infty)$ photons requires 
a powerful source of radiation with $L_{\rm h}$ between 
$10^{37}$ and $10^{38}$\es. 
For the WD's mass of $\sim 0.6$\mo\ \citep[][]{kenyon+93}, 
the maximum luminosity of the hot component could temporarily 
exceed its Eddington limit. 

To determine $L_{\rm h}$ during the 2015 outburst of AG~Peg, 
\cite{tomov+16} used, in principle, the same approach, but 
for emission lines. 
Their values of $4.5 - 6.6\times 10^{36}$\es\ are a factor 
of 19--3 below our values. This difference is caused by a 
larger opacity of the nebula in lines than in the continuum. 
For example, on July 8, 2015, their $F_{{\rm H}\beta} = 
36.3\times 10^{-11}$\ecs\ corresponds to the emission measure 
\begin{equation}
  \textsl{EM} = 4\pi d^2 \frac{F_{{\rm H}\beta}}
                {h\nu_{\beta}\,\alpha({\rm H}\beta)}
              = 3.9\times 10^{59}\,{\rm cm}^{-3},
\label{eq:eml}
\end{equation}
that is a factor of $\sim$18 smaller than our value of 
$\sim 7.0\times 10^{60}\,{\rm cm}^{-3}$ determined by 
modelling the SED. Thus, using the line fluxes gives 
significantly lower values of $L_{\rm h}$, because 
$L_{\rm h} \propto \textsl{EM}$. 

\subsection{The hot component radius}
\label{dis:rh}
During the outburst, the effective radius of the WD's 
pseudophotosphere, $R_{\rm h}$, varied between $\sim 0.09$ 
and $\sim 0.15$\ro. On average, it was a factor of 2--3 
larger than during quiescent phase. Maximum values were 
reached at/around maxima of the star's brightness. According 
to the adopted optically thick wind in nova outbursts 
(see Sect.~\ref{sec:mdot}), the optically thick/thin interface 
of the wind represents the WD's pseudophotosphere. As a consequence, 
the variations of its radius are caused by the variations 
in the mass-loss rate, $\dot M_{\rm h}$. A correlation between 
$R_{\rm h}$ and $\dot M_{\rm h}$ from the \ha\ flux 
(Fig.~\ref{fig:lrt}) supports the plausibility of the wind model. 

\subsection{The type of the 2015 outburst}
\label{dis:type}
Evolution of parameters $L_{\rm h}$, $R_{\rm h}$ and $T_{\rm h}$ 
in the \textsl{H-R} diagram is depicted in Fig.~\ref{fig:hr}. 
Optical maxima around July 1, 2015 and October 8, 2015 in 
the \textsl{LC} correspond to maxima in $L_{\rm h}$ and 
$R_{\rm h}$. During both flares, $T_{\rm h}$ of the WD's 
photosphere was very high; in the first one, significantly 
higher than during quiescence. Throughout the whole outburst, 
the optical brightening was governed exclusively by the nebular 
component of radiation. This represents a rare case of outbursts 
produced by symbiotic binaries. \cite{sk05b} classified them as 
2nd-type, and described them as being explained by a low orbital 
inclination of the binary, when the observer can directly see 
the hot WD (Fig.~\ref{fig:sketch}), and thus measures its very 
high $T_{\rm h}$, $L_{\rm h}$ and \textsl{EM}. 
No warm pseudophotosphere in the optical was indicated by modelling
the SED (Fig.~\ref{fig:sedopt}) -- the characteristic feature of
the 1st-type of outbursts produced by systems with a high $i$ 
\citep[see Fig.~26 of][]{sk05b}. 

\subsection{The extension of the He$^{+2}$ zone}
\label{dis:he2}

\begin{figure}
\begin{center}
\resizebox{\hsize}{!}{\includegraphics[angle=-90]{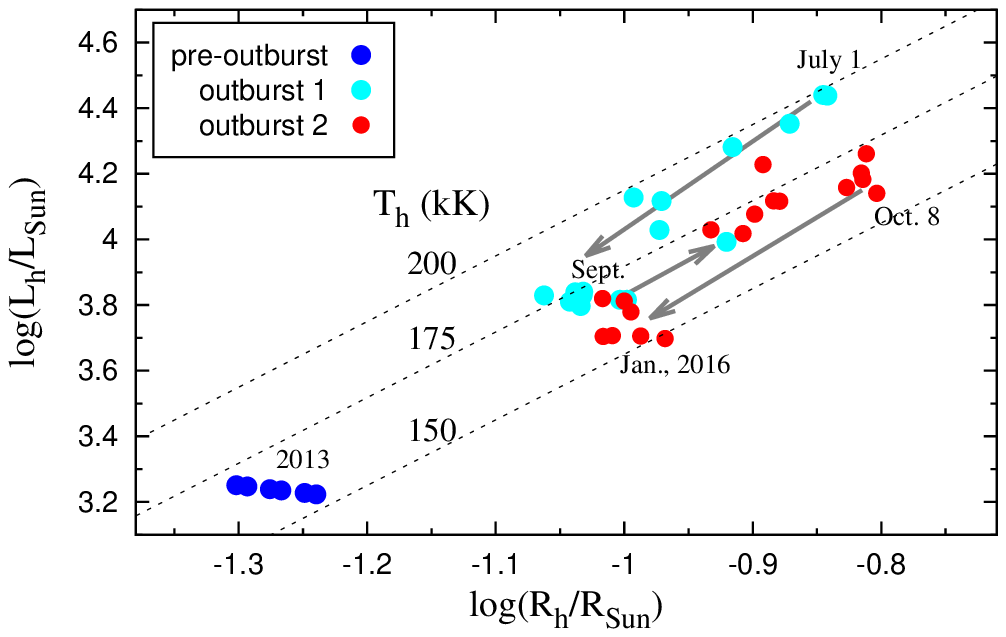}}
\end{center}
\caption[]{
Evolution of the hot component in AG~Peg in the H-R diagram. 
Its 2015 outburst was of the 2nd-type, because of a very high 
temperature of the WD pseudophotosphere (see Sect.~\ref{dis:type}). 
          }
\label{fig:hr}
\end{figure}

From the middle of August to the end of September, and from November 
25 to the end of the observing season of AG~Peg (January 2016), 
a gradual noticeable increase of $R_{\rm h}$ was observed at 
a slow gradual decline of $T_{\rm h}$ (Figs.~\ref{fig:lrt} 
and \ref{fig:hr}). The flux and the terminal velocity of the 
\ion{He}{ii}\,4686\,\AA\ line was decreasing, whereas for 
the \hb\ line these parameters did not change noticeably 
during these periods (Fig.~\ref{fig:param}). 
This implies a shrinking of the He$^{+2}$ zone, because 
the wind was always characterised with the same $v_{\infty}$, as 
indicated by a stable \hb\ and \ha\ line profile, and with the 
same $\dot M_{\rm h}$ (Fig.~\ref{fig:lrt}). During December 
2015, the broad wings in the \ion{He}{ii} profile practically 
disappeared (Fig.~\ref{fig:param}). 

\cite{leedjarv+04} found a linear relation between equivalent 
widths of the \ion{He}{ii} 4686\,\AA\ and the Raman scattered 
\ion{O}{vi} 6825\,\AA\ lines for measurements from quiescent 
phases and small brightenings. 
According to this result, the shrinkage of the He$^{+2}$ zone 
suggests a shrinkage of the O$^{+5}$ zone around the hot 
component in AG~Peg during the above mentioned periods. 
Surprisingly, the flux of the \ion{O}{vi} Raman line was 
a factor of 4.5--3.5 higher than during preceding quiescence 
(Fig.~\ref{fig:param}). 
This suggests a markedly higher Raman conversion efficiency in 
comparison with the quiescent phase. This is possible only if 
the \ion{H}{i} scattering zone covers a larger fraction of the 
O$^{+5}$ sky than in quiescence (see Sect.~\ref{dis:ram3}). 
%
%

\subsection{On the mass-loss rate}
\label{dis:mdot}
Section~\ref{sec:mdot} presents a method for determining 
the $\dot M_{\rm h}$ using the observed \textsl{EM} and \ha\ 
luminosity. The method assumes that the measured emission 
is produced by the optically thin part of the ionised wind 
from the hot component during the outburst. 
Dispersion in the $\dot M_{\rm h}$ values obtained by the \ha\ 
method is lower than that derived from the \textsl{EM} 
(see Fig.~\ref{fig:lrt}). 
This is probably a result of a larger opacity of the wind in 
the \ha\ line than in the continuum. For example, at the maximum 
of \textsl{EM} the \ha\ flux was still increasing. From this 
point of view, the adopted value of the beginning of the optically 
thin wind in \ha, $R_{\rm in} = 1.5\times R_{\rm h}$, is 
probably too rough (Sect.~\ref{sec:mdot}). 

The upper and lower limits of the $\dot M_{\rm h}$ correspond 
to maximum uncertainty of the wind outset -- at the WD's 
pseudophotosphere or at the WD's core surface. Their average is 
as high as a few $\times 10^{-6}$\,\myr, which agrees with 
values obtained independently by fitting the broad \ha\ wings 
that develop during Z~And-type outbursts of other systems 
\citep[][]{sk06}\footnote{This method is applicable only for 
systems with a high orbital inclination.}. 
The wind is variable, because the observed nebular emission 
varies during the course of the outburst. 

\subsection{The Raman-scattered \ion{O}{vi}\,6825\,\AA\ line}
\label{dis:ram}
From the properties of the Raman-scattering process and its 
induced feature observed around 6825\,\AA\ (Figs.~\ref{fig:ram1} 
and \ref{fig:ram2}; Sect.~\ref{sec:ovi}) we can infer some 
characteristics of both the neutral \ion{H}{i} and the ionised 
\ion{O}{vi} zone during the outburst. 

\subsubsection{Minima of the Raman line fluxes}
\label{dis:ram1}
Fluxes of the Raman 6825\,\AA\ line, $F_{6825}$, are in 
anticorrelation with the \textsl{EM} and thus $\dot M_{\rm h}$, 
because $\dot M_{\rm h}\propto \textsl{EM}^{1/2}$ 
(Eq.~(\ref{eq:em}), Figs.~\ref{fig:param} and \ref{fig:lrt}). 
Observing the minimum of $F_{6825}$ fluxes at a very 
high temperature rejects its usual interpretation by a cooling 
of the ionising source. Observations suggest that the 
transient weakening of the \ion{O}{vi} lines is a result 
of a significant increase of the $\dot M_{\rm h}$ 
(Fig.~\ref{fig:lrt}), which makes the O$^{+5}$ zone optically 
thick, and thus more compact. For example, the optical depth of 
the electron-scattering layer, 
$\tau_{\rm e} \propto \dot M_{\rm h}$. 
Following decrease of $\dot M_{\rm h}$ then produces a more 
transparent O$^{+5}$ zone and thus leads to an increase 
of both the $F_{1032}$ and $F_{6825}$ fluxes at similar 
conditions -- a high $T_{\rm h}$ and the geometry of \ion{H}{i} 
scatterers. 

A connection between the radius of the \ion{O}{vi} zone, 
$R_{\ion{O}{vi}}$, its $\tau_{\rm e}$, average particle density, 
$\bar{n}_{\rm e}$, and the \textsl{EM} was indicated for the 
symbiotic star AG~Dra by \cite{sk+09a}. Having independently 
determined $\tau_{\rm e}$, the authors inferred 
$R_{\ion{O}{vi}}\sim 50$\ro\ 
($\tau_{\rm e} = 0.061$, 
$\bar{n}_{\rm e} = 2.6\times 10^{10}$\cmt, and 
\textsl{EM} = $1.3\times 10^{59}$\cmt) during quiescent phase, 
and 
$R_{\ion{O}{vi}}\sim 11$\ro\ 
($\tau_{\rm e} = 0.083$, 
$\bar{n}_{\rm e} = 1.7\times 10^{11}$\cmt, and 
\textsl{EM} = $2.2\times 10^{59}$\cmt) during the transition 
from a burst, at high $T_{\rm h} = 1.5-1.6\times 10^5$\,K. 
However, the theoretical modelling is needed for a better 
understanding of the connection between $R_{\ion{O}{vi}}$, 
$\tau_{\rm e}$, $\bar{n}_{\rm e}$ and \textsl{EM} in the 
variable wind from active hot components in symbiotic binaries. 

\subsubsection{Broadening of the Raman line}
\label{dis:ram2}
Energy conservation of the Raman $1032\rightarrow 6825$\,\AA\ 
conversion implies the broadening of the scattered line with 
a factor of $(\lambda_{\rm Ram}/\lambda_{1032})^2 \sim 44$ 
\citep[][]{nussb+89}. 
It is given by the broadening of the original \ion{O}{vi} 
line, which reflects the kinematic of emitting material in the 
vicinity of the hot WD, and by the relative 
motions between the source of the original \ion{O}{vi} photons 
and the scattering \ion{H}{i} atoms. 
Firstly, the observed increase of the Raman 6825\,\AA\ line 
width during the outburst was in part caused by a broadening 
of the direct \ion{O}{vi} 1032\,\AA\ line as suggested by the 
broadening of \ion{He}{ii} and \ion{H}{i} lines in the optical. 
Secondly, a significant change of the ionisation structure 
of hot components in symbiotic binaries during outbursts 
will produce a change in the geometry and motion of \ion{H}{i} 
atoms in the binary, and thus also in the profile of 
the Raman-scattered lines (see below). 

\subsubsection{The high flux and profile of the Raman line}
\label{dis:ram3}
\begin{figure}
\begin{center}
\resizebox{\hsize}{!}{\includegraphics[angle=-90]{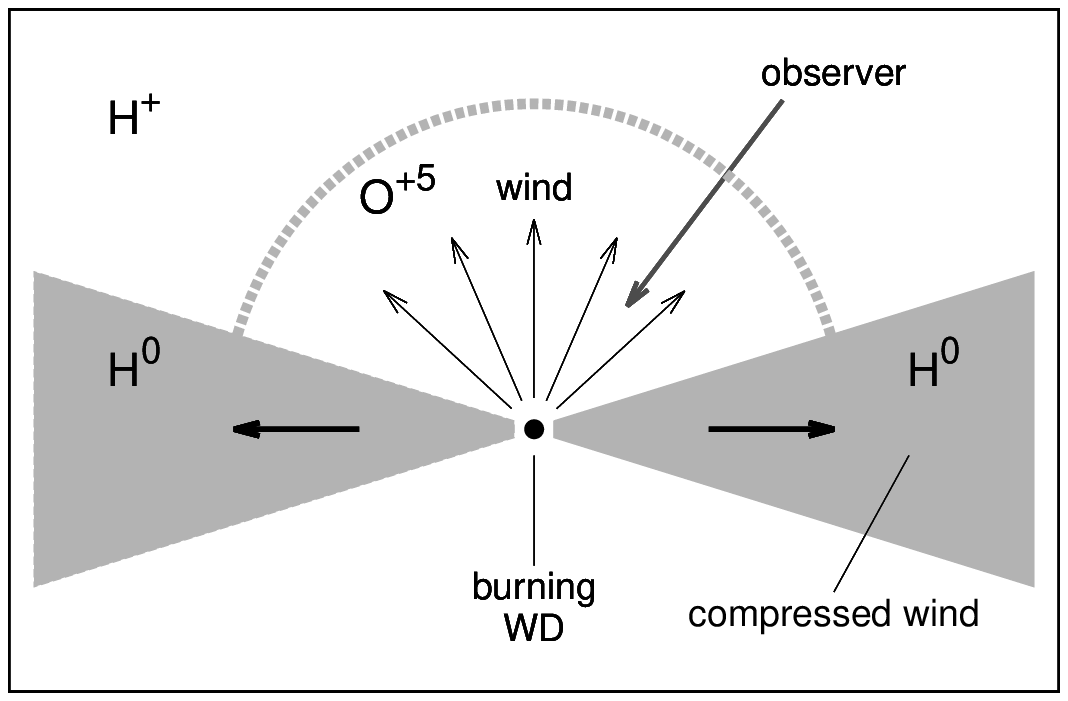}}
\end{center}
\caption[]{   
Sketch of the ionisation structure of the hot component in 
AG~Peg during its 2015 outburst (edge-on view containing 
the burning WD -- the black circle). 
The enhanced wind from the WD is ionised by its radiation. 
Due to the rotation of the WD, the wind is compressed towards 
the orbital plane, where it creates a neutral H$^{0}$ zone in 
the form of a flared disk \citep[see][]{cs12}. 
The O$^{+5}$ zone is bounded by the grey dashed line. 
Its location above/below the neutral zone significantly 
increases the efficiency of the Raman \ion{O}{vi} $\lambda 1032 
\rightarrow \lambda 6825$ scattering process 
(see Sect.~\ref{dis:ram3}). 
          }
\label{fig:sketch}
\end{figure}

The significant increase of the flux, width, and development of 
the redward-shifted component in the profile of the 
Raman-scattered \ion{O}{vi} line can be qualitatively explained 
by the presence of a neutral disk-like zone at the equatorial 
plane, which develops during active phases of symbiotic binaries 
around the hot WD \citep[see][]{sk05b,cs12}. 
The following points are relevant. 
\begin{enumerate}
\item
The cross-section of the $\lambda1032 \rightarrow \lambda6825$ 
Raman conversion, $\sigma_{\rm Ram} \sim 5\times 10^{-24}$\cmdd\ 
\citep[][]{l+l97} requires the \ion{H}{i} column density 
$n_{\rm H} \ga 10^{23}$\cmd\ along the incident \ion{O}{vi} 
photons to produce an observable effect (i.e. the optical depth 
$n_{\rm H}\sigma_{\rm Ram}\ga 1)$. The $n_{\rm H}$ throughout 
the disk-like neutral zone is of a few times $10^{22} - 10^{23}$ 
\citep[see Fig~4 of][]{cs12}, which is consistent with that 
needed for the \ion{O}{vi}\,$\lambda$1032 Raman scattering. 
Thus, the \ion{O}{vi} Raman conversion can take place mainly 
in the densest inner parts of the \ion{H}{i} zone. 
\item
Efficiency of the Raman scattering depends on the geometry of 
the scattering atoms in the binary. It is proportional to 
the so-called covering factor $C_{\rm S} = \Delta\Omega/4\pi$, 
where $\Delta\Omega$ is a solid angle, under which the initial 
O$^{+5}$ line photons, located mostly in the vicinity of the WD, 
can see the scattering region \citep[see][]{lee03}. 
During the quiescent phase, the \ion{H}{i} region has a 
cone-like shape around the giant (STB). Parameters 
of AG~Peg suggest its opening angle of $40 - 50^{\circ}$ 
(i.e. the angle between the binary axis and the asymptote to 
the ionisation boundary), which corresponds to 
$C_{\rm S} = 0.12 - 0.18$. In fact, the $C_{\rm S}$ is smaller, 
because the scattering region represents only part of the 
total \ion{H}{i} region, which is optically thick for Raman 
scattering \citep[see][]{s+s15}. 
During outbursts, location of the O$^{+5}$ zone above/below 
the neutral disk suggests that the covering factor can be as 
high as $\sim 0.5$ (see Fig.~\ref{fig:sketch}). This makes 
the Raman scattering more efficient during the outburst, thus 
producing a strong Raman line. This interpretation is 
consistent with the conclusion of Sect.~\ref{dis:he2} that 
the covering factor $C_{\rm S}$ was larger than during 
quiescence. 
\item
Observations of systems with a high orbital inclination often 
indicate P-Cyg profiles of hydrogen Balmer lines with the 
absorption component at $\sim 0.1\times v_{\infty}$ during 
outbursts. The neutral disk-like zone thus expands from the 
central ionising source at/around the orbital plane 
\citep[see Sect.~4.1 of][ in detail]{sk+06}. 
The centrally symmetric and expanding O$^{+5}$ wind nebular 
zone located above/below the neutral disk can cause a significant 
broadening of the Raman line during the outburst. 
The expanding disk-like zone of scatterers is likely to be 
responsible for the pronounced redward-shifted component at 
$\sim$6838\,\AA\ (see Fig.~\ref{fig:sketch}). 
However, a quantitative modelling of the Raman profile during 
outbursts is required to approve or disapprove this theory. 
\end{enumerate}
The above sketched interpretation is independently supported by 
the presence of the Raman-scattered \ion{O}{vi} lines in the 
spectrum of massive luminous B[e] star LHA~115-S~18 discovered 
by \cite{torres+12}. 
Because the B[e] supergiants may harbour dense circumstellar 
disks of neutral hydrogen and dust, the authors suggested 
that the Raman emission in the spectrum of LHA~115-S~18 could 
arise from such a disk 
illuminated by the radiation from the central hot star. 
Similarity of the Raman 6825\,\AA\ profile in AG~Peg and in 
LHA~115-S~18 is striking 
\citep[see Fig.~2 of][ and Fig.~\ref{fig:ram2} here]{torres+12}. 
%
%

\subsection{Shift and broadening of the \ion{He}{ii} 
            emission core}
\label{dis:shift}
During the outburst, we measured a blueward shift of the 
\ion{He}{ii}\,4686\,\AA\ emission core with respect to its 
reference wavelength (Fig.~\ref{fig:param}). The effect was 
indicated at the beginning of the outburst, but mainly 
between August 23 and November 13, 2015, when the bisectors 
measured at 0.1, 0.5 and 0.9$\times I_{\rm max}$ had similar 
RV (Fig.~\ref{fig:param}, third panel from the bottom). 
This reflected shifting of the whole line core. 
This effect was already indicated during outbursts of Z~And 
\citep[][]{sk+06,sk+09b}, and interpreted as a result 
of a disk-structured hot object that blocks a fraction of 
the redward-shifted emission from a relatively small He$^{+2}$ 
zone in the direction of the observer. As a result, 
the blueward shifted emission dominates the profile. 
The shift depends on the size of both the disk and 
the He$^{+2}$ zone. 

A significant shrinking of the \ion{He}{ii} zone from November 
25, 2015 to January, 2016 (Sect.~\ref{dis:he2}) with simultaneous 
shifting of the line core towards the reference wavelength 
(Fig.~\ref{fig:param}) can be interpreted as a result of 
a gradual dilution of the optically thick disk-like zone 
around the WD, signalling thus the end of the active phase. 

\subsection{The nature of the outburst}
\label{dis:nat}
The high luminosity of the hot component during the quiescent 
phase of $\sim 2200$\lo\ (Table~\ref{tab:sed}) requires 
accretion at a rate of a few times $10^{-8}$\myr\ to generate 
this luminosity by the stable hydrogen burning on the surface 
of a $0.6$\mo\ WD \citep[e.g.][]{sb07,nomoto+07}. 
The presence of the shell burning prior to the outburst was 
also suggested by \cite{ramsay+16}. 
However, the high accretion rate requires a high mass transfer 
ratio (i.e. the mass accretion rate divided by the mass-loss 
rate from the donor). A very efficient wind mass transfer was 
suggested, for example, by \cite{mp12}, for symbiotic stars 
with slow and dense winds. \cite{sc15} suggested that the high 
wind-mass-transfer efficiency in symbiotic stars can be caused 
by compression of the wind from their slowly rotating giants. 
Direct indication of wind focusing was found by \cite{shag+16} 
for the symbiotic binary SY~Mus. 

To increase the luminosity of the burning WD by a factor of 
$\sim$10, to values observed around the maximum of the 2015 
outburst, a transient increase of the accretion rate to 
$\dot{M}_{\rm acc} \sim 3\times 10^{-7}$\myr\ is needed 
to generate the luminosity due to hydrogen burning, 
\begin{equation}
  L_{\rm nucl.} = \eta X \dot{M}_{\rm acc}~
                  \sim ~8\times 10^{37} {\rm\,erg\,s^{-1}},
\label{eq:nucl}
\end{equation}
where $\eta = 6.3\times 10^{18}$\,erg\,g$^{-1}$ is the energy 
production of 1 gram of hydrogen due to the nuclear fusion of 
four protons, and $X\equiv 0.7$ is the hydrogen mass fraction 
in the accreted matter. 
The required $\dot{M}_{\rm acc}$ exceeds the stable-burning 
limit, which leads to blowing optically thick wind from the 
WD \citep[][]{hach+96}. 
The enhanced wind is ionised by the hot WD's pseudophotosphere, 
and thus converts a fraction of its stellar radiation to the 
nebular emission. The corresponding increase of the \textsl{EM} 
(Fig.~\ref{fig:sedopt}) then causes a relevant brightening 
in the \textsl{LC,} which we indicate as the outburst 
(Fig.~\ref{fig:ubvri}). 

However, the principal question remains as to the origin of this 
material. Probably, the red giant is responsible for a transient 
increase of the mass transfer ratio, which temporarily enhances 
$\dot{M}_{\rm acc}$. 
In general, \cite{bis+06} found that variations in the velocity 
of the wind from the giant can cause a disruption of the accretion 
disk, leading to the infall of a considerable amount of matter 
($\sim 10^{-7}$\mo) onto the WD surface, thus inducing the outburst. 
For Z~And, \cite{sok+06} suggested that its 2000-02 outburst 
was triggered by the influx of hydrogen-rich material from 
a dwarf-nova-like disk instability at its start that 
consequently enhanced nuclear shell burning on the WD. 

\section{Summary}
\label{sec:sum}
We analysed primarily high cadence optical spectroscopy 
and multicolour photometry obtained prior to and during 
the first Z~And-type outburst of the symbiotic nova AG~Peg 
that began in June of 2015. 
We determined the fundamental parameters $L_{\rm h}$, 
$R_{\rm h}$ and $T_{\rm h}$ of the WD's pseudophotosphere 
and its mass-loss rate, $\dot{M}_{\rm h}$, in the course of 
the outburst (Fig.~\ref{fig:lrt}). 
Monitoring the evolution of the \ion{He}{ii}\,4686\,\AA, \hb\ 
and the Raman-scattered 6825\,\AA\ line profiles allowed us 
to suggest the ionisation structure of the active hot component 
in AG~Peg (Fig.~\ref{fig:sketch}) and to outline the nature of 
its outburst. The main results of our analysis may be summarised 
as follows. 
\begin{enumerate}
\item
The nova-like outburst of AG~Peg terminated around 1997 when 
a wave-like orbitally-related variation in the \textsl{LC} 
reached its maximum amplitude and no overall change of 
the star's brightness was indicated until the 2015 eruption 
(Fig.~\ref{fig:ulc}). 
AG~Peg thus entered a quiescent phase, when the WD could start 
to accrete from the giant's wind (Fig.~\ref{fig:seduv}, 
Sects.~\ref{sec:phot} and \ref{sec:em}). 
\item
Evolution of the multicolour \textsl{LC}s covering the 2015 
active phase of AG~Peg corresponds to the outburst of the 
Z~And-type (Fig.~\ref{fig:ubvri}, Sect.~\ref{sec:phot}). 
\item
The brightening during the outburst was caused exclusively 
by the increase of the nebular component of radiation 
(Fig.~\ref{fig:sedopt}, Table~\ref{tab:sed}). During the 
outburst, the \textsl{EM} increased to a few times $10^{60}$\cmt, 
that is, by a factor of $\sim$10 with respect to values from 
the quiescent phase (Fig.~\ref{fig:param}, Sect.~\ref{sec:em}). 
\item
Simultaneous broadening of the line profiles and the increase 
of fluxes of hydrogen and \ion{He}{ii} 4686\,\AA\ lines 
(Figs.~\ref{fig:param} and \ref{fig:broad}) support the idea 
that the nebular emission was generated by the ionised wind 
from the hot component. 
\item
During the optical maximum, $T_{\rm h}$ increased to 
$\sim 2\times 10^{5}$\,K, and then gradually decreased 
to $\sim 1.5\times 10^{5}$\,K in January 2016, being 
comparable with values from the preceding quiescence 
(Figs.~\ref{fig:lrt} and \ref{fig:hr}, Sect.~\ref{sec:th}). 
\item
During the outburst, the luminosity $L_{\rm h}$ was as high 
as a few times $10^{37}$\es\ with a maximum of $\sim 10^{38}$\es, 
that is, a factor of $\sim$10 higher than during the preceding 
quiescence (Fig.~\ref{fig:lrt}, Sects.~\ref{sec:lhrh} and 
\ref{dis:lh}). 
\item
Assuming the so-called optically thick wind in nova outbursts, 
we determined $\dot{M}_{\rm h}$ to be of a few 
$\times 10^{-6}$\,\myr\ using our values of \textsl{EM} and 
\ha\ line fluxes (Fig.~\ref{fig:lrt}, Sect.~\ref{sec:mdot}). 
\item
The effective radius $R_{\rm h}$ of the WD's pseudophotosphere 
(= the optically thick/thin interface of the wind) 
increased by a factor of 2--3 with respect to values from 
quiescence. Maximum values of $\sim 0.15$\ro\ were indicated at 
the optical maxima (Sects.~\ref{sec:lhrh} and \ref{dis:rh}). 
\item
According to model SEDs, the 2015 outburst of AG~Peg belongs 
to the 2nd-type, because the hot component radiated at higher 
temperature than during quiescence (Sect.~\ref{dis:type}, 
Fig.~\ref{fig:hr}) and the optical brightening was caused 
solely by the nebular component of radiation. 
\item
Significant broadening and high fluxes of the Raman-scattered 
\ion{O}{vi} 6825\,\AA\ line (Figs.~\ref{fig:param}, 
\ref{fig:ram1} and \ref{fig:ram2}, Sect.~\ref{dis:ram3}) 
and probably also the blueward shift of the \ion{He}{ii} 
emission core (Sect.~\ref{dis:shift}) indicated indirectly 
the presence of the neutral disk-like formation encompassing 
the WD. This finding is consistent with the ionisation 
structure of hot components in symbiotic stars during active 
phases. 
\end{enumerate}
To ignite the 2015 outburst of AG~Peg, a transient increase in 
the accretion rate to $\sim 3\times 10^{-7}$\myr\ was required 
to generate the high luminosity ($\approx L_{\rm Edd}$ for 
a 0.6\mo\ WD) at the initial maximum (Sect.~\ref{dis:nat}). 
Such an accretion rate exceeds a critical value for the stable 
hydrogen burning. Under these conditions, optically thick wind 
will blow from the WD. At the high temperature of the ionising 
source, the wind converts a fraction of the WD's photospheric 
radiation into a strong nebular emission that dominates 
the optical (Fig.~\ref{fig:sedopt}). As a result, we observe 
a 2\,mag brightening in the \textsl{LC}, which is classified as 
the Z~And-type of the outburst (Sect.~\ref{dis:nat}). 

However, the above sketch of the Z~And-type outburst needs 
to be elaborated in more detail to understand its role in the 
evolution of accreting WDs in symbiotic stars towards their 
supposed final stage of SNe Ia type. 
In this respect, accurate quantities of the total energy output 
and the mass-loss rate during both quiescent phases and outbursts 
are required to find if the mass of the WD can reach 
the Chandrasekhar limiting mass. 

\begin{acknowledgements}
We thank the anonymous referee for constructive comments. 
Daniela Kor\v{c}\'akov\'a, Jan \v{C}echura, Jakub Jury\v{s}ek, 
Radek K\v{r}\'{\i}\v{c}ek and Petr Zasche are thanked for their 
assistance in acquisition of some spectra at the Ond\v{r}ejov 
observatory. Michaela Kraus is thanked for a discussion on 
the Raman lines in the spectrum of the B[e] star LHA~115-S~18. 
\textsl{HST} spectra presented in this paper were obtained from 
the Mikulski Archive for Space Telescopes (MAST). MAST is 
located at the Space Telescope Science Institute (STScI). 
STScI is operated by the Association of Universities for 
Research in Astronomy, Inc., under NASA contract NAS5-26555. 
The optical spectra presented in this paper were in part obtained 
within the {\it Astronomical Ring for Access to Spectroscopy 
(ARAS)}, an initiative promoting cooperation between professional 
and amateur astronomers in the field of spectroscopy, coordinated 
by Francois Teyssier. 
We also acknowledge the variable-star observations from the AAVSO 
International Database contributed by observers worldwide and used 
in this research.
This work was supported by the Czech Science Foundation, grants 
P209/10/0715 and GA15-02112S, 
by the RFBR grant No. 15-02-06178 and NSh-9670.2016.2, 
by the Slovak Research and Development Agency under the contract 
No. APVV-15-0458, by the Slovak Academy of Sciences grant 
VEGA No. 2/0008/17 and by the realisation of the project ITMS 
No.~26220120029, based on the supporting operational Research 
and development program financed from the European Regional 
Development Fund. 
\end{acknowledgements}
%
%

%
%
\appendix
\label{app:A}
\section{The distance to AG~Peg}
 
Comparing a synthetic spectrum to the $RIJKLM$ flux points 
of the giant in AG~Peg, \cite{sk05b} determined its effective 
temperature $T_{\rm eff} = 3600\pm 100$\,K and angular radius 
$\theta_{\rm g} = 2.4\times 10^{-9}\, (= R_{\rm g}/d)$, 
which corresponds to the flux 
$F_{\rm g}^{\rm obs} = 5.6\times 10^{-8}$\,\ecs\ 
(see Eq.~(\ref{eq:fbol})). 
Independently, \cite{boffin+14} measured the interferometric 
diameter of the giant in AG~Peg to $1.00\pm 0.04$\,mas, which 
yields the same $\theta_{\rm g} = 2.42\times 10^{-9}$ and 
$T_{\rm eff} = 3550\pm 120$\,K for the given $J$ and $K$ 
magnitudes. 
Using these parameters, elements of the spectroscopic orbit 
and considering a radius-luminosity relation, \cite{boffin+14} 
derived the parameters of the system for the orbital inclination 
$i$ = 90, 50 and 30$^{\circ}$ (see their Table~3). 
For the rotational velocity of $v\sin(i) = 8.5\pm 1.5$\kms, mass 
of the WD of $0.46\pm 0.1$\mo\ and the assumption of synchronism 
they preferred solutions for $i = 90^{\circ}$, which correspond 
to the distance $d = 910 - 1220$\,pc. 

However, $i \ll 90^{\circ}$, because there is no Rayleigh 
scattering attenuation of the continuum around Ly-${\alpha}$ 
at $\varphi \sim 0$ (see e.g. \textsl{IUE} spectra SWP03830, 
SWP52386 and/or the \textsl{HST} spectrum here, taken at 
$\varphi = 0.99$), which is always measured for eclipsing 
symbiotic stars during quiescent phases 
\citep[see][ for EG~And, SY~Mus, RW~Hya and BF~Cyg]{v91,dumm+99,pl96}. 
A constraint for $i$ can be obtained from a similarity 
of the ionisation structure of symbiotic stars during quiescent 
phases (see Fig.~11 of \cite{kenyon+93} for AG~Peg and 
Fig.~7 of \cite{fc+88} for Z~And). 
For Z~And, \cite{skshag12} measured a weak attenuation of the 
far-UV continuum by Rayleigh scattering at $\varphi = 0.96$, 
which restricts its $i$ to 59$\pm 3^{\circ}$. This suggests 
that $i$ of AG~Peg can be even lower than this value, because 
no Rayleigh attenuation is indicated. Therefore, 
Boffin et al. solutions for $i = 50^{\circ}$ that correspond 
to $d = 650 - 910$\,pc and $R_{\rm g} = 71.11 - 99.01$\ro\ are 
more appropriate than those for $i = 90^{\circ}$. In addition, 
fundamental $L, R, T$ parameters of the giant in AG~Peg 
satisfy statistical relations between radii, effective 
temperatures and spectral types for M giants better than those 
given by solutions for $i = 90^{\circ}$ \citep[e.g.][]{belle+99}. 
Accordingly, for the purpose of this paper, we adopted the 
distance to AG~Peg $d = 800$\,pc as originally estimated by 
\cite{kenyon+93}. 
%
\section{Comparison of black body and atmosphere models for 
         the hot component}
%
%
\begin{figure}
\begin{center}
\resizebox{\hsize}{!}{\includegraphics[angle=-90]{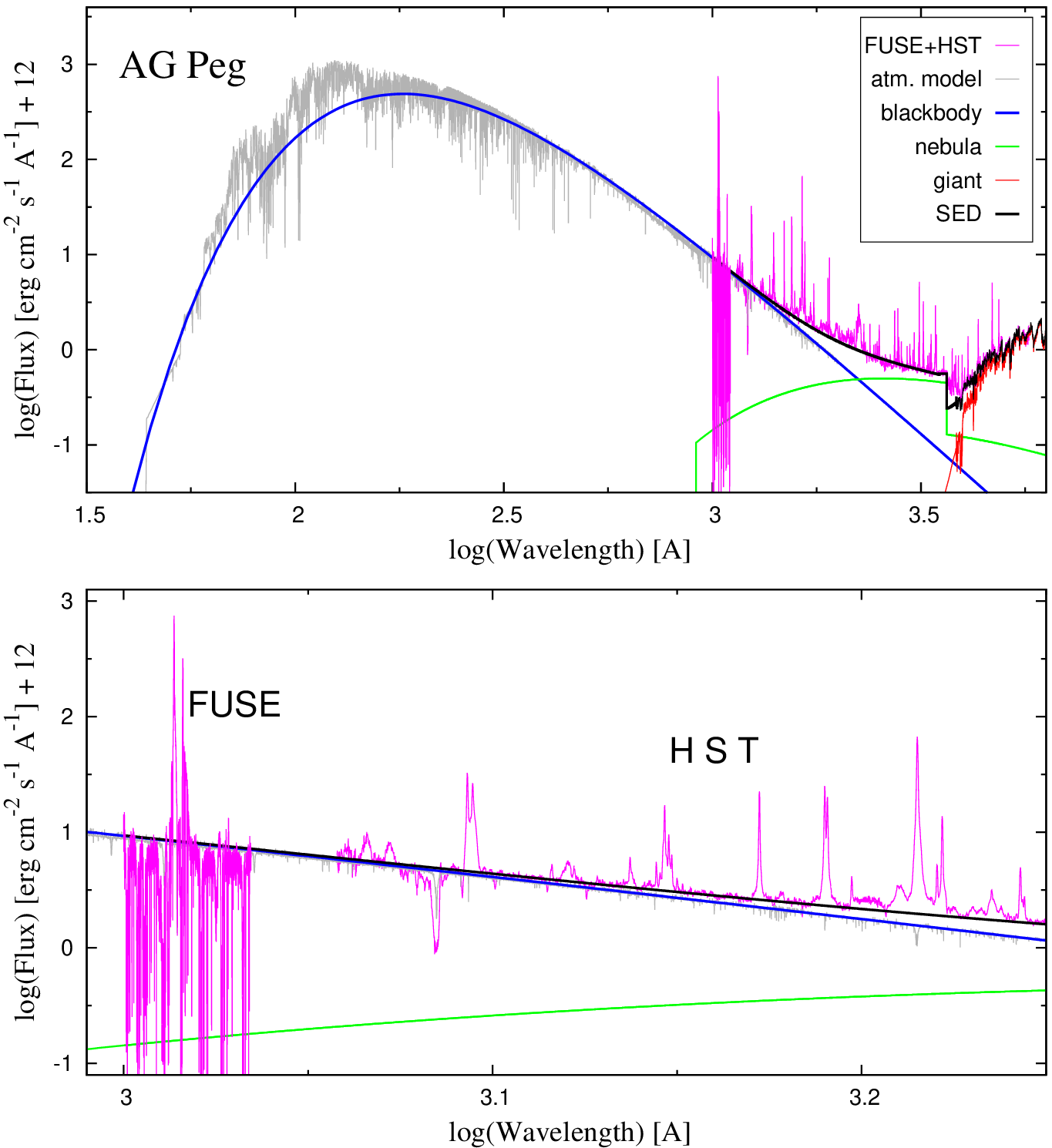}}
\end{center}
\caption[]{ 
Top: Comparison of the black body and atmosphere model for 
160000\,K scaled to the model SED from December 12, 1996 
(see Fig.~\ref{fig:seduv}). 
Bottom: A detail of the top panel between 980 and 1780\,\AA. 
Spectra were dereddened with $E_{\rm B-V} = 0.1$ (see the text). 
          }
\label{fig:atm}
\end{figure}
\cite{sk15} elaborated multiwavelength modelling of the global 
X-ray/IR SED of supersoft X-ray sources. He found that physical 
parameters of the overall SED do not basically depend on the model 
used. A black body or an atmosphere model yields a similar 
luminosity for given temperature (see Sect.~4.4 there). 

Top panel of Fig.~\ref{fig:atm} shows a comparison of the black body 
and atmosphere models for $T_{\rm eff} \equiv T_{\rm BB} = 160000$\,K 
scaled to the far-UV fluxes of the \textsl{HST} spectrum from 
December 12, 1996 (see Fig.~\ref{fig:seduv}). We used a publicly 
available NLTE atmosphere model described by \cite{rauch03} 
with the H-Ni solar abundances and $\log(g)$ = 6 (solid grey line). 
Similar profiles of the global SED for both models suggests 
also a similar quantity of the fluxes of ionising photons. 
Our illustration in Fig.~\ref{fig:atm} corresponds to 
$Q(\nu_0,\infty)_{\rm atm.} = 1.57\times 10^{47}$ 
and 
$Q(\nu_0,\infty)_{\rm BB} = 1.48\times 10^{47}$\,s$^{-1}$. 
This justifies our assumption that the hot component radiates 
as a black body, which we used in Sects.~\ref{sec:th} and 
\ref{sec:lhrh} to estimate fundamental parameters of the hot 
WD in AG~Peg. 

Bottom panel of Fig.~\ref{fig:atm} shows the \textsl{FUSE} LiF1A 
channel spectrum (1000--1082\,\AA) from June 5, 2003, dereddened 
by using the far-UV extrapolation of the \cite{c+89} extinction 
curve. Its continuum fits well the model SED at the far-UV as 
given by the \textsl{HST} spectrum from December 12, 1996. 
We note that both spectra were obtained at similar orbital phases. 
This demonstrates the reliability of the extrapolation of the 
extinction curve to $\sim 1000$\,\AA\ for small values 
of $E_{\rm B-V}$. 
However, fluxes dereddened with $E_{\rm B-V} = 0.3$ are a factor 
of 1.5--2 above the predicted model in the wavelength range 
of 1000--1100\,\AA\ \citep[see Sect.~2.2. of][ for Z~And]{sk+06}. 
%
%
\section{Analytical form of the \textsl{EM} for 
         the spherically symmetric $\beta$-law wind}

Emission measure of the completely ionised hydrogen in the volume 
$V$ is defined as, 
\begin{equation}
   \textsl{EM} = \int_{V}\!\! n_{+}n_{\rm e}\,{\rm d}V .
\label{app:em1}
\end{equation}
For the spherically symmetric ionised wind, whose particle 
density, $n(r)$, satisfies the mass continuity equation, 
\begin{equation}
   \dot M_{\rm h} = 4\pi\,r^2\,\mu m_{\rm H}\,n(r)\,v(r), 
\label{app:conity}
\end{equation}
and the particle velocity, $v(r)$, obeys the $\beta$-law wind 
(\ref{eq:betalaw}), the \textsl{EM} can be expressed as 
\begin{equation}
  \textsl{EM} = A \int_{R_{\rm in}}^{R_{\rm out}}\!\!
                  \frac{dr}{r^{2}(1-bR_0/r)^{2\beta}} .
\label{app:em2}
\end{equation} 
$r$ is the radial distance from the WD centre, 
the stellar wind begins at $r = R_0$, $R_{\rm in}$ and 
$R_{\rm out}$ are the inner and outer radius of the ionized 
volume, respectively, and 
\begin{equation}
  A = \frac{4\pi}{(4\pi\mu m_{\rm H})^2}
      \Big(\frac{\dot M_{\rm h}}{v_{\infty}}\Big)^{2}.
\label{app:Asub}
\end{equation}
Using substitutions $\sin^2(x) = bR_0/r$ and then $\cos(x) = y$, 
allow us to rewrite Eq.~(\ref{app:em2}) in the analytic form, 
\begin{equation}
 \textsl{EM} = \frac{A}{bR_0(1-2\beta)}
        \Big[\Big(1-\frac{bR_0}{R_{\rm out}}\Big)^{1-2\beta} -
        \Big(1-\frac{bR_0}{R_{\rm in}}\Big)^{1-2\beta}\Big]. 
\label{app:em3}
\end{equation}
For $R_{\rm out} = \infty$, Eq.~(\ref{app:em3}) gives 
Eq.~(\ref{eq:em}) that we used to determine $\dot M_{\rm h}$ 
from the observed \textsl{EM}. 
For $\beta = 0.5$, 
\begin{equation}
 \textsl{EM} = \frac{A}{bR_0}
               \Big[\ln\Big(1-\frac{bR_0}{R_{\rm out}}\Big) - 
               \ln\Big(1-\frac{bR_0}{R_{\rm in}}\Big)\Big]. 
\label{app:em4}
\end{equation} 
%
%
\section{Table of the used line fluxes}
%
%
\begin{table*}
\caption[]{(= {\bf Table 6}) 
Dereddened fluxes for the \ion{He}{ii}\,4686\,\AA, \hb, \ha\ 
and Raman scattered \ion{O}{vi}\,6825\,\AA\ lines in units of 
10$^{-11}$\ecs. $F_{\lambda}^{\rm obs}$ are dereddened observed fluxes, 
whereas $F_{\lambda}^{\rm tot}$ are corrected for the central absorption 
($F_{\lambda}^{\rm tot} = F_{\lambda}^{\rm obs} + F_{\lambda}^{\rm abs}$, 
Sect.~\ref{sec:he2h}). 
}
\begin{center}
\begin{tabular}{cccccccc}
\hline
\hline
Date                      &
Julian date               & 
$F_{4686}$                & 
$F_{{\rm H}\beta}^{\rm obs}$  &
$F_{{\rm H}\beta}^{\rm tot}$  &
$F_{{\rm H}\alpha}^{\rm obs}$ &
$F_{{\rm H}\alpha}^{\rm tot}$ &
$F_{\rm Raman}$           \\
yyyy/mm/dd.ddd  &
JD -- 2\,440000 &
                &
                &
                &
                &
                &
                \\
\hline
          \multicolumn{8}{c}{Quiescent phase} \\
\hline
2006/10/18.926&  54027.426&   -- &   -- &   -- &     47.1&  52.5 &   1.09 \\
2013/08/11.900&  56516.400&  5.24&  11.9&  13.5&     54.4&  66.4 &   1.21 \\
2013/08/29.945&  56534.445&  -   &   -- &   -- &     60.8&  68.6 &   1.34 \\
2013/09/04.843&  56540.343&  -   &   -- &   -- &     65.8&  75.1 &   1.33 \\
2013/09/04.855&  56540.355&  5.14&  11.9&  13.7&      -- &   --  &    --  \\
2013/09/27.978&  56563.478&  5.65&  11.9&  13.5&      -- &   --  &    --  \\
2013/10/07.004&  56572.504&  -   &   -- &   -- &     61.1&  68.8 &   1.27 \\
2013/10/24.874&  56590.374&  6.22&  12.6&  14.2&      -- &   --  &    --  \\
2013/10/26.793&  56592.293&  6.12&  11.5&  13.0&      -- &   --  &    --  \\
2013/12/17.716&  56644.216&  7.20&  12.7&  14.8&      -- &   --  &    --  \\
\hline
          \multicolumn{8}{c}{Active phase} \\
\hline
2015/06/27.227&  57200.727&   -- &    -- &   -- &     140&  150 &   1.76 \\
2015/07/01.011&  57204.511&  32.2&   37.1&  43.9&     128&  152 &   0.33 \\
2015/07/09.999&  57213.499&  40.9&   51.9&  58.8&     sat&  sat &   1.19 \\
2015/07/17.004&  57220.504&  48.8&   58.4&  68.0&     sat&  sat &   2.53 \\
2015/07/29.942&  57233.442&  35.9&   45.5&  50.9&     195&  217 &   3.71 \\
2015/08/04.971&  57239.471&  41.1&   60.0&  70.5&     sat&  sat &   3.91 \\
2015/08/11.006&  57245.506&   -- &    -- &   -- &     sat&  sat &   4.45 \\
2015/08/13.896&  57248.396&   -- &    -- &   -- &     190&  216 &   4.37 \\
2015/08/21.896&  57256.396&  27.1&   43.4&  52.0&     192&  219 &   4.39 \\
2015/08/21.953&  57256.453&   -- &    -- &   -- &     201&  224 &   4.46 \\
2015/08/23.002&  57257.502&  32.3&   50.2&  60.7&     182&  197 &   5.04 \\
2015/08/23.983&  57258.483&  31.1&   48.8&  59.5&     189&  210 &   5.20 \\
2015/08/25.039&  57259.539&  28.6&   46.1&  55.5&     176&  197 &   5.21 \\
2015/08/28.927&  57263.427&  27.5&   39.1&  48.4&     186&  208 &   5.26 \\
2015/09/06.883&  57272.383&  25.5&   41.1&  51.7&     186&  207 &   5.04 \\
2015/09/10.852&  57276.352&  25.0&   37.9&  48.8&     173&  205 &   4.81 \\
2015/09/19.875&  57285.375&  23.8&   40.2&  51.9&     207&  249 &   4.78 \\
2015/09/25.837&  57291.337&  23.0&   41.0&  50.6&     204&  237 &   4.57 \\
2015/10/01.825&  57297.325&  24.2&   42.9&  52.2&     208&  239 &   4.38 \\
2015/10/08.849&  57304.349&  28.1&   54.0&  68.9&     290&  334 &   4.37 \\
2015/10/11.779&  57307.279&  26.9&   45.2&  57.6&     255&  295 &   1.76 \\
2015/10/12.131&  57307.631&   -- &    -- &   -- &     245&  283 &   1.57 \\
2015/10/13.777&  57309.277&  26.6&   46.7&  59.2&     255&  301 &   0.47 \\
2015/10/20.830&  57316.330&  31.4&   50.8&  61.8&     289&  346 &   0.65 \\
2015/10/25.747&  57321.247&  37.0&   52.9&  60.3&     246&  284 &   1.13 \\
2015/10/27.780&  57323.280&  35.8&   67.0&  79.8&     271&  309 &   1.61 \\
2015/11/01.771&  57328.271&  29.0&   48.5&  57.9&     260&  305 &   1.07 \\
2015/11/01.807&  57328.307&  27.2&   47.6&  52.9&     242&  295 &   1.35 \\
2015/11/01.817&  57328.317&   -- &    -- &   -- &     248&  294 &   1.16 \\
2015/11/01.840&  57328.340&  27.3&   47.2&  54.0&     -- &   -- &   1.16 \\
2015/11/13.749&  57340.249&  37.2&   67.6&  74.4&     278&  325 &   2.50 \\
2015/11/25.801&  57352.301&  37.6&   69.5&  72.8&     274&  304 &   3.02 \\
2015/12/07.805&  57364.305&  32.9&   65.3&  72.9&     274&  308 &   3.11 \\
2015/12/12.803&  57369.303&  30.8&   62.8&  73.3&     260&  290 &   3.54 \\
2015/12/23.735&  57380.235&  27.3&   61.1&  68.5&     285&  317 &   3.00 \\
2015/12/29.761&  57386.261&  22.1&   50.5&  56.2&     245&  274 &   3.28 \\
2016/01/08.720&  57396.220&  25.8&   58.7&  71.0&     285&  323 &   3.57 \\
2016/01/13.723&  57401.223&  21.9&   54.8&  64.8&     254&  284 &   4.03 \\
\hline
\end{tabular}
\end{center}
%
\label{tab:fluxes}
\end{table*}
\end{document}